\documentclass[12pt,a4paper]{article}
\usepackage[longnamesfirst, authoryear, round]{natbib}

\let\counterwithin\relax
\usepackage[T1]{fontenc}
\usepackage[utf8]{inputenc}
\usepackage[width=1\textwidth, labelfont=bf]{caption}
\usepackage{enumitem}
\usepackage{booktabs}
\usepackage[all]{nowidow} 
\usepackage{amsmath}
\usepackage{amsthm} 
\usepackage{dsfont}
\usepackage{amsfonts}
\usepackage{tikz} 
\usepackage{calc}
\usepackage{siunitx}
\newcommand{\widesim}[2][1.5]{
  \mathrel{\overset{#2}{\scalebox{#1}[1]{$\sim$}}}
}

\usepackage{tikz-qtree} 
\usepackage{pgfplots}
\pgfplotsset{compat = 1.14}
\usepackage{mathtools,amssymb}
\usepackage{xcolor}
\usepackage{cooltooltips}
\usepackage{MnSymbol}
\usepackage{graphicx}
\usepackage{microtype}
\newcommand{\bigCI}{\mathrel{\text{\scalebox{1.07}{$\perp\mkern-10mu\perp$}}}} 
\usepackage{subcaption}
\usepackage[flushleft]{threeparttable}
\usepackage{multirow}

\usepackage{placeins} 
\usepackage{indentfirst} 
\usepackage{lscape}
\usepackage{rotating}
\usepackage[pdfborderstyle={/S/U/W 1},breaklinks,colorlinks]{hyperref}
\AtBeginDocument{%
  \hypersetup{
    citecolor=blue,
    urlcolor =blue,
    menucolor = black,
    linkcolor = black}}
    
\usepackage{tocloft}

\usepackage{chngcntr} 
\counterwithin{figure}{section}
\counterwithin{table}{section}

\usepackage{tcolorbox}
\tcbset{colframe=black, colback=white,size=fbox}

\usepackage{algpseudocode}
\usepackage[linesnumbered,ruled]{algorithm2e}

\usepackage[left=2.8cm,right=2.8cm,top=3.5cm,bottom=3.5cm]{geometry}

\title{CATE meets ML \thanks{\scriptsize Financial support of the European Union's Horizon 2020 research and innovation program ``FIN-TECH: A Financial supervision and Technology compliance training programme'' under the grant agreement No 825215 (Topic: ICT-35-2018, Type of action: CSA), the European Cooperation in Science \& Technology COST Action grant CA19130 - Fintech and Artificial Intelligence in Finance - Towards a transparent financial industry and the Deutsche Forschungsgemeinschaft's IRTG 1792 grant is gratefully acknowledged.}\\ {\normalsize The Conditional Average Treatment Effect and Machine Learning }}

\date{\today}

\author{Daniel Jacob \\ \href{mailto:daniel.jacob@hu-berlin.de}{daniel.jacob@hu-berlin.de}}

\begin{document}


\maketitle

\begin{abstract}

\begin{footnotesize}
For treatment effects - one of the core issues in modern econometric analysis - prediction and estimation are two sides of the same coin.
As it turns out, machine learning methods are the tool for generalized prediction models. Combined with econometric theory, they allow us to estimate  a personalized treatment effect - the conditional average treatment effect (CATE). In this tutorial, we give an overview of novel methods, explain them in detail, and apply them via Quantlets in real data applications. We study the effect that microcredit availability has on the amount of money borrowed and if 401(k) pension plan eligibility has an impact on net financial assets, as two empirical examples. The presented toolbox of methods contains meta-learners, like the DR, R-, T- and X-learner, and methods that are specially designed to estimate the CATE like causal BART and generalized random forest. In an additional simulation study we further compare the methods to observe patterns and similarities.
 \end{footnotesize}

\end{abstract}

\newpage

\section{Introduction}
\begin{figure}[ht!]
\begin{center}
\includegraphics[width=0.8\textwidth]{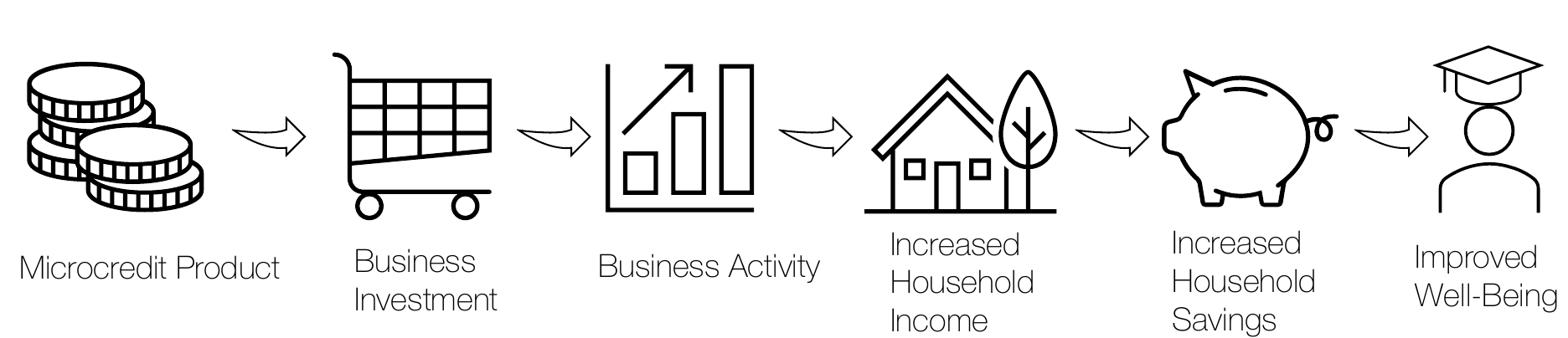}
\caption{The theory of microcredits. }
\label{fig:microcredit}
\end{center}
\end{figure}

Estimation and prediction of treatment effects are important tasks for every economist and financial econometrician since treatment effects are often the basis for policy and business decisions. As an illustration, let us look at an idea of microcredits, dating back to Muhammad Yunus, a Nobel Price winner, who discovered in 1976 that very small loans could make a disproportional difference to a poor person. Microcredits work as shown in Figure \ref{fig:microcredit}. They can increase investments since such credit is easy to get and pay back. Business activity is hence more flexible and could be improved. Increasing gains from a business could increase the household income and further allow for more savings which can be invested in, for example, education.

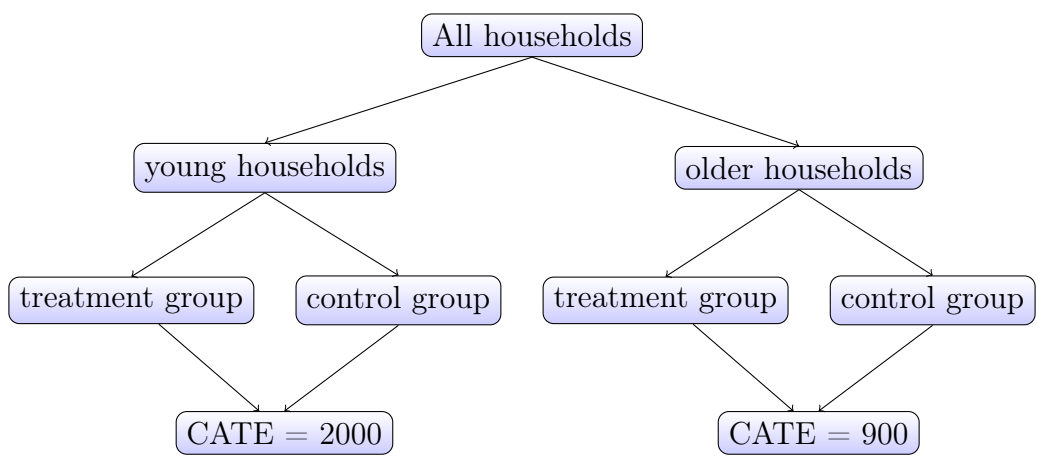
\begin{figure}[ht!]
\begin{center}
\begin{tikzpicture}[edge from parent/.append style={->},sibling distance=10em,level 2/.style= {level distance = 50pt, sibling distance=100pt},
level 1/.style= {level distance = 50pt, sibling distance=200pt},level 3/.style= {level distance = 50pt, sibling distance=200pt},
  every node/.style = {shape=rectangle, rounded corners,
    draw, align=center,
    top color=white, bottom color=blue!20}]]
  \node (root) {All households}
    child { node {young households} 
    child { node {treatment group} }
      child { node {control group}
      child[xshift=-1.5cm]  {node (cate1) {CATE = 2000}} } }
    child { node {older households}
      child { node {treatment group}}
      child { node {control group}
       child[xshift=-1.5cm]  {node (cate2) {CATE = 900}} } };
      \draw[->] (root-1-1) -- (cate1);
      \draw[->] (root-2-1) -- (cate2);
\end{tikzpicture}
\end{center}
\caption{CATE example for microcredits}
\label{Fig:CATE_emp}
\end{figure}

This specific example was recently applied by \cite{crepon2015estimating} who studied the setting where certain villages in Morocco get access to microcredit (the treatment group) while others don't (the control group). As economists, one is interested in the effect that microcredit availability has on the amount of loans which could be an indicator of how demanded such microcredits are. Since we observe certain characteristics for each household, we can condition on such observed variables to see if there is heterogeneity in the effect from microcredit. Figure \ref{Fig:CATE_emp} shows an example of what we want to do. The goal is to find subgroups based on characteristics where we believe that the treatment effect is different. As an example, we can partition the households by age and compare young vs. older household members in terms of their effect on microcredit. In both subgroups, we need to make sure that we observe people that are treated and others that did not receive treatment. We can estimate the average treatment effect (ATE) for the young household members, for example, by taking the difference of their mean outcome given treatment status. We repeat this for the subgroup of older households. Recent methods to estimate the ATE using nonparametric methods on the whole sample include target maximum likelihood estimation (TMLE) \citep{van2010targeted} and double machine learning \citep{chernozhukov2018double}. If the data has many covariates (let us say it has high-dimensionality) and if we don't know which specific subgroup we should focus on, as is the case here, we can use methods that are presented in this tutorial. These methods estimate a treatment effect for each observation based on their covariates, the conditional (on covariates) average treatment effect (CATE). In a further step, we can then look at the heterogeneity and try to link characteristics that are drivers for different treatment effects.

The high-dimensionality of a dataset does not necessarily mean that one has more covariates than observations by default. However, if we are unsure about the structural form, we could include interaction and quadratic terms, and soon the number of dimensions increases. For example, if we have 1000 observations and 30 covariates, then by only including quadratic interactions the amount of covariates increases to 495. Including up to cubic terms leads to a dimension of 5455. If we further assume that only a few covariates are dependent on the outcome and the treatment (often called the approximate sparsity assumption), the task transfers into a selection problem where standard parametric models are limited and we might want to use machine learning (ML) methods. The reason why this is the case is either that we have more covariates than observations or that the functional forms are complex and we don't know which interaction terms to include in a linear model.  

Machine learning is not easily defined. It contains many algorithms with the main focus of prediction (regression), classification, and grouping tasks like clustering. While clustering, as a form of dimensionality reduction, only uses covariates but not outcomes with labels, we call this branch unsupervised ML. The counterpart is called supervised ML. Supervised ML, in general, uses a set of covariates to predict an observed outcome. When talking about prediction we mean the following: Construct an estimator $\hat{\mu}(x)$ of $ \mathsf{E}[Y|X_i=x]$ using $Y$ and a set of covariates from some training set and predict the values of $Y$ from an independent test set. The goal is to minimize deviations between the true outcome and predicted outcomes from the test set. Note that this is in contrast to the term forecasting. The only assumption so far is that the observations are independent and that the joint distribution of $X$ and $Y$ in the training set is the same as that of the test set. To achieve the goodness of fit (for example minimize the average of the mean-squared error), in an independent test set many alternative models are estimated and the model that maximizes a criterion is selected. We will talk about cross-validation - a concept for model selection - later. The key is that the functional form is mostly determined as a function of the data. Regularization together with systematic model selection may be the main advantages of ML methods. When we talk about ML in this tutorial, we mean supervised machine learning models that are used to make predictions. For a detailed discussion about machine learning in economics see \cite{mullainathan2017machine} and \cite{atheyimpact}. 

How do we get from prediction to causal inference?  A simple, pure prediction approach to get the CATE is to estimate two conditional mean functions, one for the treated observations and one for the non-treated (the control group). For each observation, we can predict the outcome under treatment and control by plugging each observation into both functions. Taking the difference between the two outcomes results in the CATE. Mapping the support of $X$ on $Y$ is a classic regression task for which machine learning methods are well suited to find generalizable predictive patterns. Since we are only interested in getting a good prediction of the conditional mean, we do not need to know the underlying structural form of this function which enables vanilla ML methods to be sufficient. We call such functions, where the parameters are not of immediate interest, a nuisance function. While the above example of estimating the CATE is quite simple and intuitive, we will see that there are more efficient or automated methods to estimate heterogeneous treatment effects. We will also see that while prediction models are easy to evaluate, causal parameters are not. This is mainly since the objective is different. In prediction, we can optimize a goodness of fit criterion since we observe the true outcome. The causal parameter, however, is never observed in any dataset. As in econometrics, we need to carefully design methods that aim to estimate such parameters of interest, apply statistical theory and expand the set of assumptions to interpret a parameter as causal.

This tutorial is structured as follows. First, we provide an overview of the potential outcome framework and state the necessary assumptions to interpret our parameter of interest as a causal parameter. We then explain different methods that we consider, methods that are very flexible in the choice of the ML algorithm, and methods that are developed to estimate the CATE, mostly relying on tree-based algorithms. As in classical ML, we make use of sample splitting to limit overfitting and allowing for less restrictive assumptions on the nuisance functions. We cover explanations on why and how to do sample splitting and cross-validation. Next, we investigate two empirical datasets, the microcredit example, and the 401(k) pension plan survey. Last, we include a simulation study where we generate the true treatment effect. This allows us to directly compare all different methods in terms of accuracy.  
Whenever possible we provide and link to Quantlets \href{https://github.com/QuantLet/Meta_learner-for-Causal-ML}{\includegraphics[height=3mm]{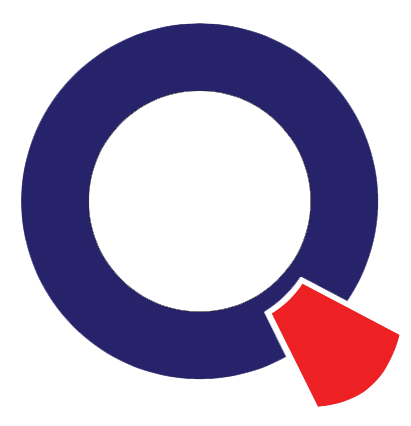}} that are ready-to-use code snippets to implement the discussed methods (the Quantlets are all written in \texttt{R}). The files are not only a replication code for the empirical analysis and the simulation study but contain functions to implement novel methods that aim to estimate the CATE directly. During this tutorial, we will use the terms model, method, and algorithm interchangeably. 

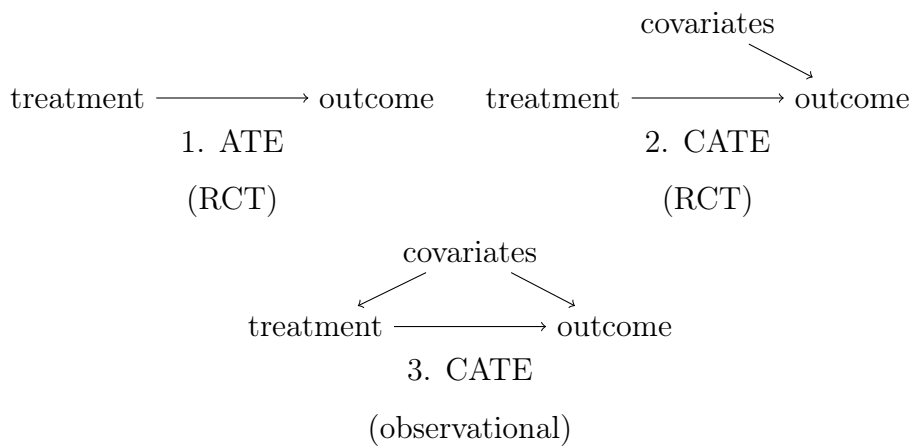
\begin{figure}[ht]
\begin{center}

 \begin{tikzpicture}[]
        \draw[->] (0,0) node[left] (D) {treatment} -- (2,0) node[right] (Y) {outcome}; 
        \path (D) -- coordinate (middle) (Y);
        \node[below of=middle, align=center] (ATEe) {1. ATE  \\(RCT)};
    \end{tikzpicture}%
    \hskip 0.4cm
     \begin{tikzpicture}[]
        \draw[->] (0,0) node[left] (D) {treatment} -- (2,0) node[right] (Y) {outcome}; 
        \path (D) -- coordinate (middle) (Y);
        \node[above of=middle] (X) {covariates};
        \draw[->] (X) -- (Y);
        \node[below of=middle,align=center] (ATE) {2. CATE \\ (RCT)};
    \end{tikzpicture}%
    \hskip 0.4cm
    \begin{tikzpicture}[]
        \draw[->] (0,0) node[left] (D) {treatment} -- (2,0) node[right] (Y) {outcome}; 
        \path (D) -- coordinate (middle) (Y);
        \node[above of=middle] (X) {covariates};
        \draw[->] (X) -- (D);
        \draw[->] (X) -- (Y);
        \node[below of=middle,align=center] (ATE) {3. CATE \\ (observational)};
    \end{tikzpicture}%

\end{center}
\caption{Simple causal diagrams - from ATE to CATE.}
\label{Fig:splitting-tree}
\end{figure}

Figure \ref{Fig:splitting-tree} gives an idea of how a causal structure may look. In the first graph, only the treatment has an impact on the outcome while the second graph also includes covariates that might make the treatment effect dependent on some characteristics. The same is true for the third graph but now the covariates also influence the treatment probability. We say that such a setting is from an observational study since the researcher has no control of the treatment assignment. The first two settings can be seen as a randomized controlled trial (RCT) but only in the second and third can we hope to observe treatment heterogeneity and hence estimate the CATE.

\section{Methods}
\setcounter{equation}{0}

Let us start with an introduction of the potential outcome framework for which we use the following notations: Each observation has two potential outcomes, $Y^1$ and $Y^0$ of which we only observe one, namely the former if someone was treated or the latter if not. The binary treatment indicator is $D \in \{0;1\}$ and we denote observed covariates $X \in \mathbb{R}$. To interpret the estimated parameter as a causal relationship, the following assumptions are needed; see, for example, \cite{rubin1980randomization}:  

\vskip 1.0cm

1. Conditional independence ( or conditional ignorability/exogeneity or conditional unconfoundedness):

\begin{align*}
\left(Y_{i}^{1}, Y_{i}^{0}\right) \bigCI D_{i}|X_{i}.
\end{align*}	

2. Stable Unit Treatment Value Assumption (SUTVA) (or counterfactual consistency):

\begin{align*}
Y_i = Y_i^0 + D_i (Y_i^1-Y_i^0).
\end{align*}

3. Overlap Assumption (or common support or positivity):

\begin{gather}
\forall x \in supp(X_i), \quad 0 < P(D_i=1|X_i=x) < 1, \nonumber \\
P(D_i=1|X_i=x) \overset{\mathrm{def}}{=} e(x).\label{equ:prop}
\end{gather}

4. Exogeneity of covariates:

\begin{align*}
X_i^1 = X_i^0.
\end{align*}

Assumption 1 together with Assumption 4 is very natural since they state that the treatment assignment is independent of the two potential outcomes and that the covariates are not affected by the treatment. Assumption 2 ensures that there is no interference, no spillover effects, and no hidden variation between treated and non-treated observations. Assumption 3 states that no subpopulation defined by $X_i = x$ is entirely located in the treatment or control group, hence the treatment probability needs to be bounded away from zero and one. Equation (\ref{equ:prop}) is called the propensity score. 

Now we define the conditional expectation of the outcome for the treatment or control group as 

\begin{align*}
\mu_d(x) = \mathsf{E}[Y_i | X_i = x, D_i = d] \quad with \quad D \in \{0,1\}.
\end{align*}
If we don't use any subscript, we refer to this function as the general conditional expectation.

Our parameter of interest is the CATE ($\tau(x)$), which is formally defined as: 

\begin{align}
\tau(x)=\mathsf{E}\left[Y_i^1-Y_i^0 \mid X_{i}=x\right]=\mu_{1}(x)-\mu_{0}(x).
\end{align}

Equation \ref{equ:cate_derivation} shows how the two conditional mean functions can represent the two potential outcomes and hence, by taking the difference, lead to the CATE. 
\begin{align}
\tau(x) &=\mu_{1}(x)-\mu_{0}(x)  \nonumber \\ 
 &=\mathsf{E}\left[Y_{i} \mid D_{i}=1,  {X}_{i}= x\right]-\mathsf{E}\left[Y_{i} \mid D_{i}=0,  {X}_{i}= x\right] \nonumber \\
  &=\mathsf{E}\left[Y_{i}^{1} \mid D_{i}=1,  {X}_{i}= x\right]-\mathsf{E}\left[Y_{i}^{0} \mid D_{i}=0,  {X}_{i}= x\right]\nonumber \\ 
   &=\mathsf{E}\left[Y_{i}^{1} \mid  {X}_{i}= {x}\right]-\mathsf{E}\left[Y_{i}^{0} \mid  {X}_{i}=x\right] \nonumber \\ 
&=\mathsf{E}\left[Y_{i}^{1}-Y_{i}^{0} \mid  {X}_{i}= {x}\right] \label{equ:cate_derivation}
 \end{align}

This estimator is of special interest in many areas like medicine or policy actions since it tells us if there are differences in the treatment effect in the population and how big these differences are. It could be, for example, that the average treatment effect of a policy is $+2$, containing half of the people with a treatment effect of $+6$ and the other half of $-2$. Instead of treating everyone, we should only treat people that have a positive effect from the policy (if positive means better). If this is not possible, let us say due to legal or ethical reasons, the policy should not be implemented at all. The CATE will tell us the exact distribution of the effects and, at best, allows us to identify subgroups. To estimate the CATE, we are not primarily interested in the coefficient from regressing $X$ on $Y$, nor are we interested in the coefficients from the propensity score model. What we want instead is to have a good approximation of the function and hence good estimates from e.g. $\mu_{1}(x)$ and $\mu_{0}(x)$. This is why ML methods are well suited for the job. 

When reviewing recently proposed methods for the estimation of the CATE, we can categorize them into two groups. The first group contains methods that are built on off-the-shelf machine learning methods (such as the lasso, random forest (RF), Bayesian additive regression trees (BART), boosting methods, or neural networks). Since the base learners are not designed to estimate the CATE directly, the literature calls them meta-learners or generic ML algorithms. The second group of methods alters existing machine learning methods in a way that they can be used to estimate the CATE directly (examples are causal boosting by \cite{powers2018methods}, causal forest by \cite{athey2019generalized} or Bayesian regression tree models for causal inference by \cite{hahn2020}). See \cite{kuenzel2019meta} for a comparison between meta-learners like the S-, T-, and X-learner as well as the causal forest in a simulation study. \cite{knaus2018machine} compare meta-learners like inverse probability weighting (IPW) estimator, doubly-robust (DR), modified covariate method (MCM), R-learner, and different versions of the causal forest in an empirical Monte Carlo study while \cite{nie2017quasi} compare their R-learner with the S-, T-, X- and U-learner as well as causal boosting. Regarding the base learners (the ML methods), \cite{kuenzel2019meta} use a random forest and BART. \cite{knaus2018machine} use RF and lasso while \cite{nie2017quasi} use boosting and the lasso for the estimation of the nuisance functions and the treatment effects. In high dimension, the use of machine learning methods, such as boosting or random forests to estimate the propensity score, works quite well as \cite{mccaffrey2004propensity} and \cite{wyss2014role} show. The estimation of probabilities given a large set of covariates is nothing less than a prediction problem in where ML methods are superior. In Table \ref{tab:methods_quantlet} we list popular methods by category, including links to the Quantlets. The references refer to recent papers that use these methods and provide theoretical properties.

\begin{table}[ht]
\centering
\caption{Methods to estimate CATE }
\label{tab:methods_quantlet}
\begin{tabular}{llll}
\hline \hline
Category                             & Method                             & Reference  & Quantlet\\ \hline
\multirow[t]{5}{*}{\textbf{Meta-Learner}}       & DR-learner                         &  \cite{kennedy2020optimal}   &\href{https://github.com/QuantLet/Meta_learner-for-Causal-ML/tree/main/DR-learner}{\includegraphics[height=5mm]{qletlogo_tr.png}$_{DR}$}       \\
												& IPW-learner                          &            \cite{horvitz1952generalization} &\href{https://github.com/QuantLet/Meta_learner-for-Causal-ML/tree/main/IPW-learner}{\includegraphics[height=5mm]{qletlogo_tr.png}$_{IPW}$} \\
                                     & R-learner                          &            \cite{nie2017quasi} &\href{https://github.com/QuantLet/Meta_learner-for-Causal-ML/tree/main/R-learner}{\includegraphics[height=5mm]{qletlogo_tr.png}$_{R}$} \\
                                     &S-learner							  &				\cite{hill2011bayesian} & \href{https://github.com/QuantLet/Meta_learner-for-Causal-ML/tree/main/S-learner}{\includegraphics[height=5mm]{qletlogo_tr.png}$_{S}$} \\
                                     & T-learner                          &            \cite{hansotia2002incremental} &\href{https://github.com/QuantLet/Meta_learner-for-Causal-ML/tree/main/T-learner}{\includegraphics[height=5mm]{qletlogo_tr.png}$_{T}$} \\
                                     & X-learner                          &            \cite{kuenzel2019meta} &\href{https://github.com/QuantLet/Meta_learner-for-Causal-ML/tree/main/X-learner}{\includegraphics[height=5mm]{qletlogo_tr.png}$_{X}$}  \\ \hline
\multirow[t]{3}{*}{\begin{tabular}[t]{@{}l@{}}\textbf{Modified ML}\\ \textbf{Methods}\end{tabular}}&
 Causal BART &     \cite{hahn2020}  &\href{https://github.com/QuantLet/Meta_learner-for-Causal-ML/tree/main/Causal-BART}{\includegraphics[height=5mm]{qletlogo_tr.png}$_{CBART}$} \\
&Causal Boosting                    &    \cite{powers2018methods}   &\href{https://github.com/QuantLet/Meta_learner-for-Causal-ML/tree/main/Causal-Boosting}{\includegraphics[height=5mm]{qletlogo_tr.png}$_{CB}$}   \\
&Causal Forest                      &  \cite{athey2019generalized}   &\href{https://github.com/QuantLet/Meta_learner-for-Causal-ML/tree/main/GRF}{\includegraphics[height=5mm]{qletlogo_tr.png}$_{GRF}$}       \\

                                     \hline \hline
\end{tabular}
\end{table}

\subsection{Meta-Learners}

In the following, we briefly describe the considered meta-learners. We follow the definition of meta-learners by \cite{kuenzel2019meta} and describe them as methods to estimate the CATE using ML methods that are built for regression or classification tasks only. The S- and T-learner, for example, can use any vanilla ML method to predict the conditional outcome. The prediction models can then be used to estimate the conditional average treatment effects. The second class of methods uses additional information from the propensity score. They contain the DR- and X-learner. Again, the conditional mean functions, as well as the propensity score, can be estimated using a broad range of ML methods. The aforementioned methods are also called transformed outcome methods. The idea is to generate a pseudo-outcome using the estimated nuisance functions in the first step. This can be seen as an approximation of the conditional average treatment effect. The pseudo-outcome, which we show in Table \ref{tab:learners}, is an unbiased estimate of the CATE given that the nuisance parameters are known (e.g. if we would know the true propensity score). In a second step, the pseudo-outcome is mapped on the covariates to get the final estimate and to make predictions on new observations. The reason for prediction is that the observed data after a treatment assignment includes the outcome, covariates, and the treatment assignment variable. If we want to classify new observations, we only observe the covariates. Hence we need a model that maps the covariates on the estimated treatment effect. The mapping also serves as a smoother since it could be the case that some pseudo-outcome values are quite extreme (e.g. if the propensity score estimate is very low or high). The last method that we examine in this category is the R-learner. However, it does not generate a pseudo-outcome in the classical sense as it needs algorithms that can modify the loss function. Still, many ML methods can be used which is why we include the R-learner into the category of meta-learners. Currently, R-packages are available for the R-, S-, T-, U-, and X-learner (\texttt{install\_github("xnie/rlearner")}) and the M-, S-, T-, and X-learner (\texttt{install\_github("soerenkuenzel/causalToolbox")}). Causal analysis via the potential outcome framework and causal graph theory for Python can be found in \cite{dowhy}. For heterogeneous treatment effect analysis via machine learning in Python see \cite{econml}. The list of methods above is not a complete list of methods in this subject. For example, we do not talk about methods for instrumental variables, multiple-treatment, difference-and-difference methods, or regression discontinuity designs.  We also note that there are other methods to estimate treatment effects in cross-sectional settings. For example, one of the first methods developed to control for confounding bias is the inverse probability weighting (IPW) estimator by \cite{horvitz1952generalization}. In Algorithm \ref{pseudo:IPW} we show how to implement this method. Some methods use neural networks to estimate heterogeneous treatment effects. See, for example, the recent work by \cite{farrell2021deep} who use a deep neural network for semiparametric inference and develop nonasymptotic high probability bounds.

\subsubsection{ Single- (S-learner) and two-model learner (T-learner):}

Let us first start with a very simple and intuitive method, the T-learner. It is a two-step approach where the conditional mean functions $\mu_1(x) = \mathsf{E}[Y^1 |X_i=x]$ and $\mu_0(x) = \mathsf{E}[Y^0 |X_i=x]$ are estimated separately with any generic machine learning algorithm. The difference between the two functions results in the CATE as shown in Table \ref{tab:learners} and as seen in equation \ref{equ:cate_derivation}. One problem with the T-learner is that it aims to minimize the mean squared error for each separate function rather than the mean squared error of the treatment effect. By splitting the sample into two groups there is only information on one group. This might be problematic if the two functions shrink different covariates which are important in both groups. This is especially the case in an RCT.  See, for example, \cite{kuenzel2019meta, kennedy2020optimal} for settings when the T-learner is not the optimal choice.  
An alternative is to model only one function and include the treatment assignment into this function. This approach is called the S-learner. See for example, \cite{hill2011bayesian} and \cite{foster2011subgroup} for early examples of proposing the S-learner. Algorithm \ref{pseudo:s} in the Appendix describes how to implement the S-learner while algorithm \ref{pseudo:t} shows the implementation for the T-learner.

\begin{algorithm}[ht!] 
\footnotesize
\SetKwInOut{Input}{Input}
\Input{$Z_i = \{Y_i,D_i,X_i\}_{i \in N}$} 
Split sample $Z$ into $K$ random subsets \\
\For{k in \{1, \ldots,K\}}{
\textbf{assign} Sample $S_a = Z \cupdot S_k$ and $S_k$ \\
 \textbf{regress} $Y^0_{i}=\hat{\mu}_{0}\left(X^0_{i}\right)+\hat{U}^0_{i},$ with $i \in S_a | D=0$ \\
 \textbf{regress} $Y^1_{i}=\hat{\mu}_{1}\left(X^1_{i}\right)+\hat{U}^1_{i},$ with $i \in S_a | D=1$ \\
 \hskip 1.0cm 	\textbf{estimate} $\hat{Y}^0_{i}=\hat{\mu}_{0}\left(X_{i}\right)$, with $i \in S_k$ \\
 \hskip 1.0cm 	\textbf{estimate} $\hat{Y}^1_{i}=\hat{\mu}_{1}\left(X_{i}\right)$, with $i \in S_k$ \\
 \textbf{create} $\hat{\tau}_k(X_i) = \hat{\mu}_1(X_i) - \hat{\mu}_0(X_i)$ \\
 }
  \textbf{combine} $\hat{\tau}(X_i) = \{\hat{\tau}_1,\hat{\tau}_k, \ldots,\hat{\tau}_K$\} \\
  \caption{T-learner}
  \label{pseudo:t}
\end{algorithm}
\noindent

\subsubsection{Doubly-Robust learner (DR-learner):}

A more efficient method than the T-learner can be the DR-learner. It builds on the T-learner and adds a version of the inverse probability weighting (IPW) scheme on the residuals of both regression functions \{$Y^d - \hat{\mu}_d(x)$\}. We can think of it as combining two different models and hence avoid drawbacks like the minimization goal from the T-learner and a potentially high variance from an IPW model when some propensity scores are small.
The doubly-robust learner takes its name from a double robustness property which states that the estimator remains consistent if either the propensity score model or the conditional outcome model is correctly specified \citep{lunceford2004stratification}. The DR-learner creates a pseudo-outcome which is an unbiased estimator for the CATE:

\begin{align}
\hat{\psi}_{DR} = \hat{\mu}_1(x) - \hat{\mu}_0(x)+\dfrac{D\left\{Y-\hat{\mu}_{1}\left(x \right)\right\}}{\hat{e}\left(x\right)} -\dfrac{\left(1-D\right)\left\{Y-\hat{\mu}_{0}\left(x \right)\right\}}{\left(1-\hat{e}\left(x\right)\right)}. \label{equ:dr-full}
\end{align}

Equation \ref{equ:dr-proof} in the Appendix shows the double-robustness property by rewriting equation \ref{equ:dr-full}. Whenever the propensity score or the conditional mean function is correctly specified the doubly-robust estimator converges to $\textsf{E}[Y^1] - \textsf{E}[Y^0]$.

The first part in equation \ref{equ:dr-full} can be seen as a regression adjustment parameter (this is the difference between the two conditional mean functions). The second part, which makes use of the propensity score, can be seen as an inverse probability weighting estimator applied to the residual from the conditional mean functions. The DR-learner is very flexible in the choice of the ML method. Estimating the nuisance parameters we can use any ML method. Since the pseudo-outcome is already an unbiased estimator for the CATE the loss-function in the final regression task is to minimize the mean squared error. This allows using the same range of ML methods as in the first step. 
Recently, this estimator has gained popularity to estimate the CATE, especially in high-dimensional settings. See, for example, the work by \cite{fan2019estimation}. Most recently, \cite{kennedy2020optimal} find that for estimating the CATE, the finite-sample error-bound from the DR-learner at most deviates from an oracle error rate by the product of the mean squared error of the propensity score and the conditional mean estimator. As can be seen from equation \ref{equ:dr-full}, extreme propensity score estimates can lead to large pseudo-outcome estimates. Hence it is necessary to look at the distribution of the propensity score and, if necessary, apply methods to make the overlap assumption more realistic.

\vskip 0.5cm
\begin{algorithm}[ht!] \label{pseudo:dr}
 \SetKwInOut{Input}{Input}
 \footnotesize
\Input{$Z_i = \{Y_i,D_i,X_i\}_{i \in N}$} 
Split sample $Z$ into $K$ random subsets \\
\For{k in \{1, \ldots,K\}}{
\textbf{assign} Sample $S_a = Z \cupdot S_k$ and $S_k$ \\
\textbf{regress} $D_{i}=\hat{e}\left(X_{i}\right)+\hat{V}_{i},$ with $i \in S_a $ \\
 \textbf{regress} $Y^0_{i}=\hat{\mu}_{0}\left(X^0_{i}\right)+\hat{U}^0_{i},$ with $i \in S_a | D=0$ \\
 \textbf{regress} $Y^1_{i}=\hat{\mu}_{1}\left(X^1_{i}\right)+\hat{U}^1_{i},$ with $i \in S_a | D=1$ \\
 \hskip 1.0cm 	\textbf{estimate} $\hat{D}_{i}=\hat{e}\left(X_{i}\right)$, with $i \in S_k$ \\
 \hskip 1.0cm 	\textbf{estimate} $\hat{Y}^0_{i}=\hat{\mu}_{0}\left(X_{i}\right)$, with $i \in S_k$ \\
 \hskip 1.0cm 	\textbf{estimate} $\hat{Y}^1_{i}=\hat{\mu}_{1}\left(X_{i}\right)$, with $i \in S_k$ \\
 \textbf{create} $\hat{\psi}_{DR,k} = \hat{\mu}_1(x) - \hat{\mu}_0(x)+\dfrac{D\left\{Y-\hat{\mu}_{1}\left(x \right)\right\}}{\hat{e}\left(x\right)}-\dfrac{\left(1-D\right)\left\{Y-\hat{\mu}_{0}\left(x \right)\right\}}{\left(1-\hat{e}\left(x\right)\right)}$ \\
 
 \textbf{store} $\hat{\psi}_{DR,k}$ for $i \in S_k$ \\
}
 \textbf{combine} $\hat{\psi}_{DR} = \{\hat{\psi}_{DR,1},\hat{\psi}_{DR,k}, \ldots,\hat{\psi}_{DR,K}\}$ \\
 \textit{Cross-fitting:}\\
 \For{oob in (1:2)}{
 \textbf{if} oob = 1: $S_{oob} = Z_i$ with $i \in \{1,...,N/2\}$ and $S_{train} = Z_i \cupdot S_{oob}$ \\
 \textbf{if} oob = 2: $S_{train} = Z_i$ with $i \in \{1,...,N/2\}$ and $S_{oob} = Z_i\cupdot S_{in}$ \\
 \For{l in 1:5}{
 \textbf{split} $S_{train}$ in $\{S_1,S_2,\ldots,S_5\}$ \\
 \textbf{regress} $\hat{\psi}_{i} = \hat{t}_{DR}(X_i) + W_i$, for $i \in S_{l}$ \\
  \hskip 1.0cm\textbf{estimate} $\tilde{\tau}_{l}(X_i) = \hat{t}_{DR}(X_i)$, with $i \in S_{oob}$ \\
  }
  \textbf{average} $\hat{\tau}_{oob}(X_i) = \textsf{E}[\tilde{\tau}(X_i)]$\\ 
  }
 \textbf{row bind} $\hat{\tau}(X_i) = \{\hat{\tau}_1,\hat{\tau}_2\}$ \\
  \caption{DR-learner}
\end{algorithm}

\subsubsection{R-learner:}

The R-learner makes use of the idea of orthogonalization to cancel out any selection bias that may arise in observational studies from observed covariates. Here, the residuals from the regression of $Y$ on $X$ are regressed on the residuals from the regression of $D$ on $X$ and weighted by the squared residuals, $\{D-\hat{e}(x)\}^2$. This is similar to the double machine learning approach from \cite{chernozhukov2018double} where their estimator of interest is the ATE.  \cite{nie2017quasi} develop a general class of two-step algorithms for the estimation of the CATE. The R-learner, as from residualized and as an homage to Peter M. Robinson, makes explicit use of machine learning methods. 
Achieving Neyman orthogonality using a residuals-on-residuals (or debiasing) approach has a long history in econometrics (see the Frisch–Waugh–Lovell theorem from the 1930s for linear regression) and mainly builds on the work by \cite{robinson1988root} who replaces the linear parts by non-parametric kernel regression. The CATE from the R-learner is obtained by the following minimization task: 

\begin{align}
\begin{split}
\hat{\tau}(\cdot)=\operatorname{argmin}_{\tau}\left\{\frac{1}{n} \sum_{i=1}^{n}\left[\left\{Y_{i}-\hat{\mu}^{(-i)}\left(X_{i}\right)\right\}\right.\right. \\
\left.\left.-\left\{D_{i}-\hat{e}^{(-i)}\left(X_{i}\right)\right\} \tau\left(X_{i}\right)\right]^{2}+\Lambda_{n}\{\tau(\cdot)\}\right\}.
\end{split}
\end{align}

The superscript $(-i)$ indicates the sample splitting. The conditional mean functions are trained without the $i$-th observations and evaluated only for $i$. We will explain certain sample splitting procedures later. The term $\Lambda_{n}\{\tau(\cdot)\}$ can be interpreted as a regularizer on the complexity of the $\tau(\cdot)$ function. In practice, this regularization term could be explicitly given as in penalized regression or implicitly introduced, e.g., as provided by a carefully designed deep neural network. The main difference to the pseudo-outcome estimators (the DR- and X-learner) is that the R-learner needs to alter the loss-function of the ML method. Even if $\psi_R$ with weights equal to $1$ as an estimator for $\tau(x)$ it can suffer from high variance if the nuisance functions are not known. The variance is mainly caused by the propensity score since $\{D-\hat{e}(x)\}$ is in the denominator. This is where the weighting comes into play. Observations that have a high variance are weighted by the squared of  $\{D-\hat{e}(x)\}$ and hence are less important. The weights for each observation directly influence the loss function (e.g. in boosting methods they manipulate the gradient). Therefore, applying the R-learner needs ML methods that have the option of altering the loss-function through weighting. The following methods have this option: lasso and ridge regression (\texttt{glmnet}), boosting (included in the DMatrix format) (\texttt{XGBoost}), neural network (\texttt{nnet}) and the random forest (\texttt{ranger}). Note that the ranger package seems to be the only implementation of weights for a random forest. The weights are applied on the whole training sample and observations with larger weights will be selected with higher probability in the bootstrap (or subsampled) samples for the trees.

\vskip 0.5cm
\begin{algorithm}[ht!] \label{pseudo:R}
\footnotesize
    \SetKwInOut{Input}{Input}
\Input{$Z_i = \{Y_i,D_i,X_i\}_{i \in N}$} 
Split sample $Z$ into $K$ random subsets \\
\For{k in \{1, \ldots,K\}}{
\textbf{assign} Sample $S_a = Z \cupdot S_k$ and $S_m = S_k$ \\
\textbf{regress} $D_{i}=\hat{e}\left(X_{i}\right)+\hat{V}_{i},$ with $i \in S_a $ \\
 \textbf{regress} $Y_{i}=\hat{\mu}\left(X_{i}\right)+\hat{U}_{i},$ with $i \in S_a $ \\
 \hskip 1.0cm 	\textbf{estimate} $\hat{D}_{i}=\hat{e}\left(X_{i}\right)$, with $i \in S_m$ \\
 \hskip 1.0cm 	\textbf{estimate} $\hat{Y}_{i}=\hat{\mu}\left(X_{i}\right)$, with $i \in S_m$ \\
 \textbf{create} $\hat{\psi}_{R} = \frac{(Y_i-\hat{\mu}(X_i)}{(D_i-\hat{e}(X_i))}$ and $w_i =  (D_i - \hat{e}(X_i))^2$\\
 \textbf{store} $\hat{\psi}_{R,k}$ and $w_{i,k}$ for $i \in S_k$ \\
}
 \textbf{combine} $\hat{\psi}_{R} = \{\hat{\psi}_{R,1},\hat{\psi}_{R,k}, \ldots,\hat{\psi}_{R,K}\}$, $w_R = \{w_1,w_k,\ldots,w_K\}$ \\
  \textit{Cross-fitting:}\\
 \For{oob in (1:2)}{
 \textbf{if} oob = 1: $S_{oob} = Z_i$ with $i \in \{1,...,N/2\}$ and $S_{train} = Z_i \cupdot S_{oob}$ \\
 \textbf{if} oob = 2: $S_{train} = Z_i$ with $i \in \{1,...,N/2\}$ and $S_{oob} = Z_i\cupdot S_{in}$ \\
 \For{l in 1:5}{
 \textbf{split} $S_{train}$ in $\{S_1,S_2,\ldots,S_5\}$ \\
 \textbf{regress} $\hat{\psi}_{i} = \hat{t}_{R}(X_i) + W_i$ and weight by $w_R$, for $i \in S_{l}$ \\
  \hskip 1.0cm\textbf{estimate} $\tilde{\tau}_{l}(X_i) = \hat{t}_{R}(X_i)$, with $i \in S_{oob}$ \\
  }
  \textbf{average} $\hat{\tau}_{oob}(X_i) = \textsf{E}[\tilde{\tau}(X_i)]$\\ 
  }
 \textbf{row bind} $\hat{\tau}(X_i) = \{\hat{\tau}_1,\hat{\tau}_2\}$ \\
  \caption{R-learner}
\end{algorithm}

\subsubsection{X-learner:}

\cite{kuenzel2019meta} propose the X-learner which estimates a treatment effect separately for the control and the treatment group. This might be especially helpful in situations where the proportion of the two groups is highly imbalanced.  
The X-learner has several steps. The first step is identical to the T-learner, namely estimating the two conditional mean functions. In the second step, we predict the counterfactual outcome using the two functions. If a person is treated and hence the observed outcome is $Y^1$ we subtract the estimated counterfactual. If a person is in the control group we use the estimated counterfactual outcome and subtract the observed outcome ($Y^0$).
This results in two imputed treatment effects:

\begin{align}
\hat{\psi}_{X}^1\overset{\mathrm{def}}{=}Y^{1}-\hat{\mu}_{0}\left(x^{1}\right) \quad \text{for} \quad D_i = 1,\\
\hat{\psi}_{X}^0\overset{\mathrm{def}}{=}\hat{\mu}_{1}\left(x^{0}\right)-Y^{0} \quad \text{for} \quad D_i = 0.
\end{align}

These imputed effects are now used in a third step to regress them individually on the covariates to obtain $\hat{\tau}_0(x)$ (the CATE for the control group) and $\hat{\tau}_1(x)$ (the CATE for the treatment group). The final estimator combines the two estimators plus some weights, $g(x)$:

\begin{align*}
\hat{\tau}(x)=g(x) \hat{\tau}_{0}(x)+\{1-g(x)\} \hat{\tau}_{1}(x).
\end{align*}

In a randomized controlled trial, the two estimates should not differ significantly. If there are confounding variables and if the support of the treatment variable given covariates differs among the treatment status we would expect the two estimates to be different. Hence, a natural weighting function for $g(x)$ could be the propensity score. 
We would use $1 -\hat{e}(x)$ for the treatment group and $\hat{e}(x)$ for the control group estimate, respectively. 

Algorithm \ref{pseudo:X} describes the procedure in detail and takes sample-splitting into account. 

\vskip 0.5cm
\begin{algorithm}[ht] 
\footnotesize
    \SetKwInOut{Input}{Input}
\Input{$Z_i = \{Y_i,D_i,X_i\}_{i \in N}$} 
Split sample $Z$ into $K$ random subsets \\
\For{k in \{1, \ldots,K\}}{
\textbf{assign} Sample $S_a = Z \cupdot S_k$ and $S_k$ \\
\textbf{regress} $D_{i}=\hat{e}\left(X_{i}\right)+\hat{V}_{i},$ with $i \in S_a $ \\
 \textbf{regress} $Y^0_{i}=\hat{\mu}_{0}\left(X^0_{i}\right)+\hat{U}^0_{i},$ with $i \in S_a | D=0$ \\
 \textbf{regress} $Y^1_{i}=\hat{\mu}_{1}\left(X^1_{i}\right)+\hat{U}^1_{i},$ with $i \in S_a | D=1$ \\
 \hskip 1.0cm 	\textbf{estimate} $\hat{D}_{i}=\hat{e}\left(X_{i}\right)$, with $i \in S_k$ \\
 \hskip 1.0cm 	\textbf{estimate} $\hat{Y}^0_{i}=\hat{\mu}_{0}\left(X_{i}\right)$, with $i \in S_k$ \\
 \hskip 1.0cm 	\textbf{estimate} $\hat{Y}^1_{i}=\hat{\mu}_{1}\left(X_{i}\right)$, with $i \in S_k$ \\
 \textbf{create} $\hat{\psi}_{X}^1\overset{\mathrm{def}}{=}Y^{1}-\hat{\mu}_{0}\left(x^{1}\right)$ for $i \in S_k$\\
 \textbf{create} $\hat{\psi}_{X}^0\overset{\mathrm{def}}{=}\hat{\mu}_{1}\left(x^{0}\right)-Y^{0})$ for $i \in S_k$ \\
 \textbf{store} $\hat{\psi}_{X}^1$, $\hat{\psi}_{X}^0$ and $\hat{e}\left(X_{i}\right)$ for $i \in S_k$ \\
}
 \textbf{combine} $\hat{\psi}_{X}^1 = \{\hat{\psi}_{X,1}^1,\hat{\psi}_{X,k}^1, \ldots,\hat{\psi}_{X,K}^1\}$, $\hat{\psi}_{X}^0 = \{\hat{\psi}_{X,1}^0,\hat{\psi}_{X,k}^0, \ldots,\hat{\psi}_{X,K}^0\}$, $\hat{e}\left(X_{i}\right)= \{\hat{e}_1\left(X_{i}\right),\hat{e}_k\left(X_{i}\right), \ldots,\hat{e}_K\left(X_{i}\right)\}$ \\
 \textbf{regress} $\hat{\psi}_{X}^1 = \hat{t}^1(X_i) + W^1_i$, with $i \in Z$ \\
  \textbf{regress} $\hat{\psi}_{X}^0 = \hat{t}^0(X_i) + W^0_i$, with $i \in Z$ \\
  \hskip 1.0cm\textbf{estimate} $\hat{\tau}^1_k(X_i) = \hat{t}^1(X_i)$, with $i \in Z$ \\
   \hskip 1.0cm\textbf{estimate} $\hat{\tau}^0_k(X_i) = \hat{t}^0(X_i)$, with $i \in Z$ \\
  \textbf{average} $\hat{\tau}_k(X_i) = \hat{e}(X_i)\hat{\tau}^0_k + (1-\hat{e}(X_i))\hat{\tau}^1_k$ \\
   \caption{X-learner}
   \label{pseudo:X}
\end{algorithm}

\subsubsection{Summary of meta-learners:}

We summarise the considered meta-learners in Table \ref{tab:learners} where $\hat{\psi}$ states the pseudo-outcome or estimator for each of the learners. 
The last column counts the number of nuisance functions needed to estimate the pseudo-outcome or estimator. In brackets, we state the total number of models needed to get the final CATE estimate. Note that the X-learner is regressed only for the treated observations and again only for the observations in the control group. This is why we need two more additional models for the final estimate.

\begin{table}[ht]
\centering
{\renewcommand{\arraystretch}{1.5}%
    \begin{threeparttable}
\caption{Summary of meta-learners}
\label{tab:learners}

\begin{tabular}{lccc}
\hline \hline
Method                 & Estimator/Pseudo-outcome & Weights ($w_i$) & \# of Models   \\  \hline

S-learner              &  $\hat{\psi}_{S} = \hat{\mu}(x,d=1) - \hat{\mu}(x,d=0)$               & 1                 &    1 (2)                      \\

T-learner              &  $\hat{\psi}_{T} = \hat{\mu}_1(x) - \hat{\mu}_0(x)$               & 1                 &    2 (3)                      \\
DR-learner             &  \begin{tabular}[c]{@{}l@{}}$\hat{\psi}_{DR} = \hat{\psi}_{T}+\dfrac{D\left\{Y-\hat{\mu}_{1}\left(x \right)\right\}}{\hat{e}\left(x\right)}$\\ \quad \quad $-\dfrac{\left(1-D\right)\left\{Y-\hat{\mu}_{0}\left(x \right)\right\}}{\left(1-\hat{e}\left(x\right)\right)}$\end{tabular}    & 1  &  3  (4) 					\\ 
R-learner              &  $\hat{\psi}_{R} = \dfrac{\{Y-\hat{\mu}\left(x\right)\}}{\{D-\hat{e}\left(x\right)\}}$      &$\left\{D-\hat{e}(x)\right\}^2$      &  2 (3)  					 \\
IPW-learner            &    $\hat{\psi}_{IPW} =\dfrac{DY}{\hat{e}(x)} - \dfrac{(1-D)Y}{(1-\hat{e}(x))} $               &       1        &  1 (2)  						\\
X-learner              &  \begin{tabular}[c]{@{}l@{}} $\hat{\psi}_{X}^1 \overset{\mathrm{def}}{=}Y^{1}-\hat{\mu}_{0}\left(x^{1}\right)$ \\
$\hat{\psi}_{X}^0\overset{\mathrm{def}}{=}\hat{\mu}_{1}\left(x^{0}\right)-Y^{0}$ \end{tabular}    & 1   &  3  (5)                     \\ \hline \hline
\end{tabular}
\begin{tablenotes}
      \small
      \item \textit{Notes:} Considered meta-learners that estimate the CATE. \# of Models counts the number of nuisance functions to estimate the pseudo-outcome. Numbers in brackets count the total number of models to train to get the final CATE estimate or to make predictions.
    \end{tablenotes}
  \end{threeparttable}
}
\end{table}

The estimators from Table \ref{tab:learners} can be represented as a weighted minimization problem which solves the following:

\begin{align*}
\min _{\tau}\left\{N^{-1} \sum_{i=1}^{N} w_{i}\left\{\hat{\psi}_{i}-\tau\left(x\right)\right\}^{2}\right\}.
\end{align*}

\subsubsection{The choice of ML algorithms for meta-learners:}

The accuracy of the CATE estimation depends on the accuracy of the nuisance functions and hence on the choice of the ML method. To minimize the dependence of the ML methods on our estimates, we do not assign specific machine learning methods for the estimation but consider a range of different popular methods. To choose which ML method to use for each nuisance function as well as for any additional functions, we use a stacking method. In such a setting, not only one ML method may be chosen but an ensemble of methods that are stacked together with different weights. We use the SuperLearner package as proposed by \cite{polley2011super}. It also enables us to choose different models for each nuisance function and setting. 
The package offers a general class of prediction methods to be considered by the ensemble. From the 42 different algorithms, we select gradient boosted trees (\texttt{xgboost}), neural network (\texttt{nnet}) and random forest (\texttt{ranger}) for our analysis. Note that the R-learner needs to include weights to minimize the R-loss in the algorithm, so we need to make sure that the ML methods we use have this possibility included. While many researchers include the lasso in their simulations or empirical analysis, we do not use this approach. The reason is that the lasso algorithm would ideally need to assume a parametric form. This means that if we believe that there are interaction and or polynomial effects from $X$ on $Y$, we would need to include such transformations. There are extensions like the adaptive lasso that expand the feature space by including such additional factors. The computation time does however increase the more features we include. These are the main reasons why we do not include the lasso in our analysis.

We use 10-fold cross-validation to estimate the performance of all machine learning models. Cross-validation is a resampling procedure used to evaluate ML models on a finite data sample. Depending on the ML model, the data can be fit perfectly and hence produce a high variance (overfitting). This is, however, on the training sample and the model can behave poorly on unseen data. Hence, we have to validate our models. We could use a part of the data for validation. Since there is never enough data, removing a part of it poses a potential for underfitting (we might lose trends in the data or important patterns). What we require instead is a method that provides enough data for training the model and also leaves enough data for validation. K-fold cross-validation does exactly that. This approach involves randomly splitting the set of observations into $K$ groups, or folds, of approximately equal size. The model is fit on folds 2 to $K$ while the first fold is used as a validation set. It is also important that any preparation of the data before fitting the model occur on the training sample that is used for cross-validation within the loop rather than on the broader data sample. This also applies to any hyperparameter tuning, for example, the number of trees, the minimum observations within a node, learning rates, or shrinkage parameters. There is no formal rule for the choice of $K$ but usually, it is set to 5 or 10. These values have been shown empirically to yield test error rate estimates that suffer neither from excessively high bias nor from very high variance. The reason is the following: The larger $K$, the smaller the difference in size between the (original) training set and the resampling subset ($K-1$ folds). As this difference decreases, the bias of the technique becomes smaller. This means that the bias is smaller for $K = 10$ than for $K = 5$. A special case of cross-validation is the leave-one-out cross-validation (LOOCV). In this case, $K$ is set to the sample size and only one observation is the validation sample. In all procedures, the $I$ resampled estimates of performance are summarized (e.g. by the mean and the standard error).

Since we apply multiple models to estimate the nuisance functions, we create a weighted average among all models. Using stacking, we can find the optimal combination of a collection of prediction algorithms or even different settings within one model. 
In other words, we build a linear model that uses the outcome variable of the validation set as the dependent variable and all different base learners as the input variables. For the random forest, we set the following tuning parameters: \texttt{n.trees=1000, min.node.size=10}.

\subsection{Modified ML Methods}
We now describe some methods that modify existing ML methods to estimate the CATE directly. In contrast to meta-learners that are flexible in the choice of the ML algorithm, these methods use a specific ML method (mostly tree-based algorithms). Packages or code in R are available for the causal forest (\texttt{grf}), the causal Boosting (\texttt{https://github.com/saberpowers/causalLearning}) and the causal BART (\texttt{install \_github("vdorie/bartCause")}. Since causal boosting is computationally expensive, we do not consider this method in our analysis.

\subsubsection{Causal Forest:}

The causal forest method, part of the generalized random forest (GRF) by \cite{athey2019generalized} builds on a random forest algorithm to find neighborhoods in the covariate space. These neighborhoods are built by recursive splitting the covariates into subgroups while the criterion to do so is based on heterogeneity in treatment effects. The idea is to find leaves where the treatment effect is constant but different from other leaves. If we know that $\tau(x)$ were constant over some neighbourhood $N(x)$, we could solve a partially linear model over $N(x)$ using the residual-on-residual approach (see e.g. \cite{robinson1988root}): 
First, we estimate $e(x)= \mathsf{E}[D_i|X_i =x]$ and second, $\mu(x)= \mathsf{E}[Y_i|X_i =x]$. We can use any non-parametric method like the
lasso, random forests, boosting methods, neural networks and others. The final step is to estimate $\tau(x)$ over the neighbourhood $N(x)$:

\begin{align}
\hat{\tau}(x)=\frac{\sum_{\left\{i: X_{i} \in \mathcal{N}(x)\right\}}\left\{Y_{i}-\hat{\mu}\left(X_{i}\right)\right\}\left\{D_{i}-\hat{e}\left(X_{i}\right)\right\}}{\sum_{\left\{i: X_{i} \in \mathcal{N}(x)\right\}}\left\{D_{i}-\hat{e}\left(X_{i}\right)\right\}^{2}}. \label{equ:R}
\end{align}

Note that this approach looks similar to the R-learner. \cite{chernozhukov2018double} showed that when using any of the aforementioned ML methods for the estimation of the nuisance functions and then use the residual-on-residual approach to estimate the average treatment effect the following regularity condition holds: 

Given that, 
\begin{align}
\mathsf{E}\left[\left\{\mu\left(X_{i}\right)-\hat{\mu}\left(X_{i}\right)\right\}^{2}\right]^{\frac{1}{2}} \ll \frac{1}{n^{1 / 4}}, \quad \mathsf{E}\left[\left\{e\left(X_{i}\right)-\hat{e}\left(X_{i}\right)\right\}^{2}\right]^{\frac{1}{2}} \ll \frac{1}{n^{1 / 4}},
\end{align}
we get a central limit theorem such that $\sqrt{n}(\hat{\tau}-\tau) \Rightarrow \mathcal{N}(0, V)$.
The treatment effect in the above setting, however, has to be constant. We can assume that with heterogeneous treatment effects, there are subgroups such that the constant effect assumption holds. The question of how to find such accurate subgroups is exactly where the (causal) random forest comes into play. To create leaves that consist of observations with the same (average) treatment effect, the splitting criterion has to rely on maximizing the heterogeneity in treatment effects between leaves (similar to maximizing the variance between the leaves). Here we use again the method from equation (\ref{equ:R}). In observational studies where self-selection into treatment is present, the first splits might not be a good representation of the treatment effect rather than differences due to confounding variables. To overcome this problem, \cite{athey2019generalized} suggest applying local-centering. This means that we use the residuals of the outcome and treatment variable as data instead of the original values. Therefore one has to train two nuisance functions beforehand to predict the conditional mean which is used to create the residuals. While machine learning methods rely on sample splitting to avoid overfitting, the causal random forest integrates this via an honesty condition. A tree is honest if, for each training sample $i$, it only uses the response $Y_i$ to estimate the within-leaf treatment effect or to decide where to place the split, but not both. 

So far we have looked at how a single tree is build and how the final treatment effect can be estimated. To extend this procedure to multiple trees, let us view a forest as a weighting function:

\begin{align}
\hat{\mu}(x)=B^{-1} \sum_{b=1}^{B} \sum_{i=1}^{n} Y_{i} \frac{1\left\{X_{i} \in L_{b}(x)\right\}}{\left|L_{b}(x)\right|}=\sum_{i=1}^{n} Y_{i} \underbrace{B^{-1} \sum_{b=1}^{B} \frac{1\left\{X_{i} \in L_{b}(x)\right\}}{\left|L_{b}(x)\right|}}_{\alpha_{i}(x)} .
\end{align}

Instead of seeing a forest as a double average over observations within a leaf and $B$ single trees, we can integrate the first sum to be a weighted average over all $X_i$ that fall into the leaf $L_b(x)$ and divide by the total number of observations within the leaf ($|L_b(x)|$). This weighted average tells us how often $Y_i$ falls into a certain leaf and hence the weight that we have to apply to control for the different proportions. The weights can be represented as $\alpha_i(x)$. We can now use these weights to weigh each observation in a generalized method of moments estimator where we apply a linear model, regressing the residuals of $D_i$ on the residuals of $Y_i$ and weigh by $\alpha_i$. This is how we get the CATE using a random forest. The algorithm is implemented in the \texttt{grf} package. See \cite{friedberg2018local} for an extension of this approach to local linear forests. 

Algorithm \ref{pseudo:CF} describes the approach to estimating the CATE for each observation using the causal forest. The results from steps 4-7 are used for local-centering, as described above. If not provided in the causal forest (step 8), the nuisance functions are estimated internally. We use the $\texttt{regression\_forest}$ function to estimate the nuisance parameters. This function uses the honest estimation which means the prediction is based on out-of-bag observations. We state the estimation explicitly since it might be the case that a different ML method is better suited in predicting the conditional mean or the propensity score (e.g a boosting method). When using different methods, we just need to make sure that the predictions (steps 6 and 7) are again based on either out-of-bag observations or a different subsample. Theoretically, if relying completely on the causal forest we do not need to split the sample at all since the honest condition applies to each step (the nuisance parameters and the estimation of the CATE). Since we use $K$ fold sample splitting for all other methods we apply the same subsamples when using the causal forest.

\begin{algorithm}[ht!]
\footnotesize
\SetKwInOut{Input}{Input}
\Input{$Z_i = \{Y_i,D_i,X_i\}_{i \in N}$} 
Split sample $Z$ into $K$ random subsets\\
\For{k in \{1, \ldots,K\}}{
\textbf{assign} Sample $S_a = Z \cupdot S_k$ and $S_k$ \\
\textbf{regress} $D_{i}=\hat{e}\left(X_{i}\right)+\hat{V}_{i},$ with $i \in S_a $ \\
 \textbf{regress} $Y_{i}=\hat{\mu}\left(X_{i}\right)+\hat{U}_{i},$ with $i \in S_a$ \\
 \hskip 1.0cm 	\textbf{estimate} $\hat{D}_{i}=\hat{e}\left(X_{i}\right)$, with $i \not\in L_b$ \\
 \hskip 1.0cm 	\textbf{estimate} $\hat{Y}_{i}=\hat{\mu}\left(X_{i}\right)$, with $i  \not\in L_b$ \\
 \textbf{apply} Causal Forest using $Y_i,D_i,X_i,\hat{\mu}(X_i), \hat{e}(X_i)$ with $i \in S_a$ \\
 \hskip 1.0cm \textbf{estimate} $\hat{\tau}_k(X_i)$ with $i \in S_k$ \\
  }
  \textbf{combine} $\hat{\tau}(X_i) = \{\hat{\tau}_1,\hat{\tau}_k, \ldots,\hat{\tau}_K$\} \\
  \caption{Causal Forest}
   \label{pseudo:CF}
\end{algorithm}
\noindent

\subsubsection{Causal Boosting:}

An alternative to random forest based causal inference is given by \cite{powers2018methods} who introduces boosted trees and causal multivariate adaptive regression splines (MARS). By iteratively fitting weak learners to the residuals of a model, an approximation of the function is build. The idea is to fit a causal tree in the style of \cite{wager2018estimation} while setting the basis function $\hat{G}(x,D)$ to zero. Now we estimate the residuals by $Y_i - \epsilon \times \hat{g}_k(X_i,D_i)$ and update $\hat{G}_k = \hat{G}_{k-1} + \epsilon \times \hat{g}_k$. $k$ defines the terminal nodes from the tree and $\epsilon$ is the learning rate parameter. After $K$ iterations we return $\hat{G}_K(x,D)$. Estimating the CATE is done by setting $D$ to 1 for the treated observations and 0 for the control-group observations, such that:

\begin{align}
\hat{\tau}(x) = \hat{G}_K(x,1) - \hat{G}_K(x,0).
\end{align}

Like in the causal forest the problem remains how to control for overfitting. Especially boosting methods are prone to overfit the data since the trees are not built independently. While a random forest would benefit from using more trees over which to average, in gradient boosting the number of trees is an important tuning parameter that needs to be controlled. In supervised ML we would ideally apply cross-validation. In our case, the parameter of interest is the CATE and we do not observe the true value for each observation. Hence, cross-validation does not apply here. Instead, we can do something like the honest approach from the causal forest. 

\cite{powers2018methods} propose to split the data into two distinct sets. The training set is used to build the causal boosting. Using the split-points and split-variables from the training set we use the covariates from the validation set, lets call it $X_v$, for validation and get new estimates based on $D_v$ and $Y_v$ for each terminal node. This procedure is done for any of the $K$ trees, using again the residuals (this time from the validation set tree) to reestimate the terminal nodes of the next causal tree. This allows estimating a validation error for each of the original $K$ models. The overall validation error for a causal boosting model is given by the differences of the CATE from the original vs. the validation trees.

\subsubsection{Causal BART:}

While (causal) boosting relies on multiplying each sequential tree by the learning rate ($\epsilon$), the idea developed by \cite{chipman2010bart} is to estimate a posterior distribution of the prediction by explicitly setting priors for the trees and ensemble structure (e.g. the depth of the tree, the probability of a new split). Using a Bayesian approach allows for a broader set of information than the point estimate from regression and classification methods. The Bayesian Additive Regression Trees (BART) approach is a combination of three methods: Using gradient boosting trees, a Bayesian framing for each individual tree, and Markov chain Monte Carlo (MCMC) sampling to do backfitting (using additive and generalized additive models for posterior sampling). \cite{hill2011bayesian} proposes to use such nonparametric Bayesian models to estimate treatment effects. Given strong ignorability, one way to estimate treatment effect is to estimate the response function $\mu(X_i,D_i)$. This function is estimated in one step instead of estimating two functions. Hence, the prior is set directly for the response surface. This approach is also called the S-learner - train one function and set $D_i$ to 1 and 0 for each observation to get estimates for both potential outcomes. \cite{hahn2020} extends the idea of using a Bayesian approach to estimate treatment effects but expresses the response surface as:

\begin{align}
\mathsf{E}\left[Y_{i} \mid {X}_{i}, D_{i}=d\right]=\mu\left\{{X}_{i}, \hat{e}\left({X}_{i}\right)\right\}+\tau\left({X}_{i}\right) D_{i},
\end{align}

where $\hat{e}(x)$ is the estimated propensity score and the functions $\mu(\cdot)$ and $\tau(\cdot)$ are independent BART priors. The inclusion of the estimated propensity score can be seen as a covariate-dependent prior to control for confounding bias. The method is specially designed to estimate the CATE from observational studies with small effect sizes and heterogeneous effects. The package we use is built on the model by \cite{hill2011bayesian} (\texttt{install\_github("vdorie/bartCause")}). A package that implements the method proposed by \cite{hahn2020} is in development (\texttt{install\_github("socket778/XBCF")}. This package is also available for Python.  Note that the causal BART produces credible intervals as a contrast to confidence intervals. They are estimated from the posterior probability function and hence rely on the prior distribution while confidence intervals are based on data only. We will only use the term confidence interval on all methods, however, we do mean credible intervals for the causal BART and (frequentists) confidence intervals for all other methods.

\begin{algorithm}[ht!]
\footnotesize
\SetKwInOut{Input}{Input}
\Input{$Z_i = \{Y_i,D_i,X_i\}_{i \in N}$} 
Split sample $Z$ into $K$ random subsets\\
\For{k in \{1, \ldots,K\}}{

 \textbf{apply} Causal BART using $Y_i,D_i,X_i$ with $i \in S_a$ \\
 \hskip 1.0cm \textbf{estimate} $\hat{\tau}_k(X_i)$ with $i \in S_k$ \\
  }
  \textbf{combine} $\hat{\tau}(X_i) = \{\hat{\tau}_1,\hat{\tau}_k, \ldots,\hat{\tau}_K$\} \\
  \caption{Causal BART}
   \label{pseudo:CBART}
\end{algorithm}
\noindent

\subsection{Sample splitting and cross-fitting} \label{sec:sample_splitting}

To aim for a consistent estimator, we need to assume certain complexity conditions on the nuisance functions. Specifically, we want them to be smooth (i.e. differentiable) and the entropy of the candidate nuisance functions to be small enough to fulfill Donsker conditions (e.g. if we assume Lipschitz parametric functions or VC classes). In high-dimensional settings (p>n) or when using ML methods that are complex or adaptive, the Donsker conditions might not hold; see, for example, \cite{robins2013new}, \cite{chernozhukov2016locally} and \cite{rotnitzky2017multiply}. As \cite{chernozhukov2018double} noticed, verification of the entropy condition is so far only available for certain classes of machine learning methods, such as lasso and post-lasso. For classes that employ cross-validation or for hybrid methods (like the SuperLearner), it is likely difficult to verify such conditions. Luckily, there is an easy solution available: sample splitting. When splitting the sample, we can use independent sets for estimating the nuisance functions and constructing the treatment estimation equation. By using different sets, we can treat the nuisance functions as fixed functions which allow avoiding conditions on the complexity. It also allows us to use any ML method such as random forest or boosting or even an ensemble of different methods. The split-sample approach to avoid smoothness conditions dates back at least to \cite{bickel1982adaptive} and was extended to also use cross-fitting by \cite{schick1986asymptotically}. 

To overcome a potential loss in efficiency, since only a subset of the data is used when estimating the CATE, cross-fitting is an increasingly popular approach to combine ML methods with semi-parametric estimation problems; see, for example, \cite{chernozhukov2018double}, \cite{newey2018cross} and \cite{athey2017efficient}. We note that there are two definitions of cross-fitting. First, it is defined in the context of estimating the CATE for all observations. For example, we split the data into two folds, subset A and M. We use fold A to train the nuisance functions and then estimate the parameter of interest using subset M. Now we switch the roles of the sets, using subset M for training and subset A for estimation. As a result, we get estimates of the CATE for all observations. 

The second definition is more in the spirit of averaging CATE estimates obtained from different partitions that are used for the nuisance parameter estimation. For example, let us say we have again the data as above but also an independent test set. Now we can use the procedure as before. First, we train the nuisance function on subset A and predict on subset B to get the pseudo-outcomes. We again train a regression function based on B but predict the CATE using the test set. Now we reverse the roles of A and B and get a second prediction of the CATE for the test set. The two results are now averaged to get the final estimate. In this tutorial, we will combine the two definitions of cross-fitting. First, we estimate the CATE on all observations through reversing roles of samples. Second, we use cross-fitting as an averaging tool over $K$ folds. When referring to cross-fitting we mainly mean the latter definition. 

\begin{figure}[ht]
\begin{center}
\includegraphics[width=0.8\textwidth]{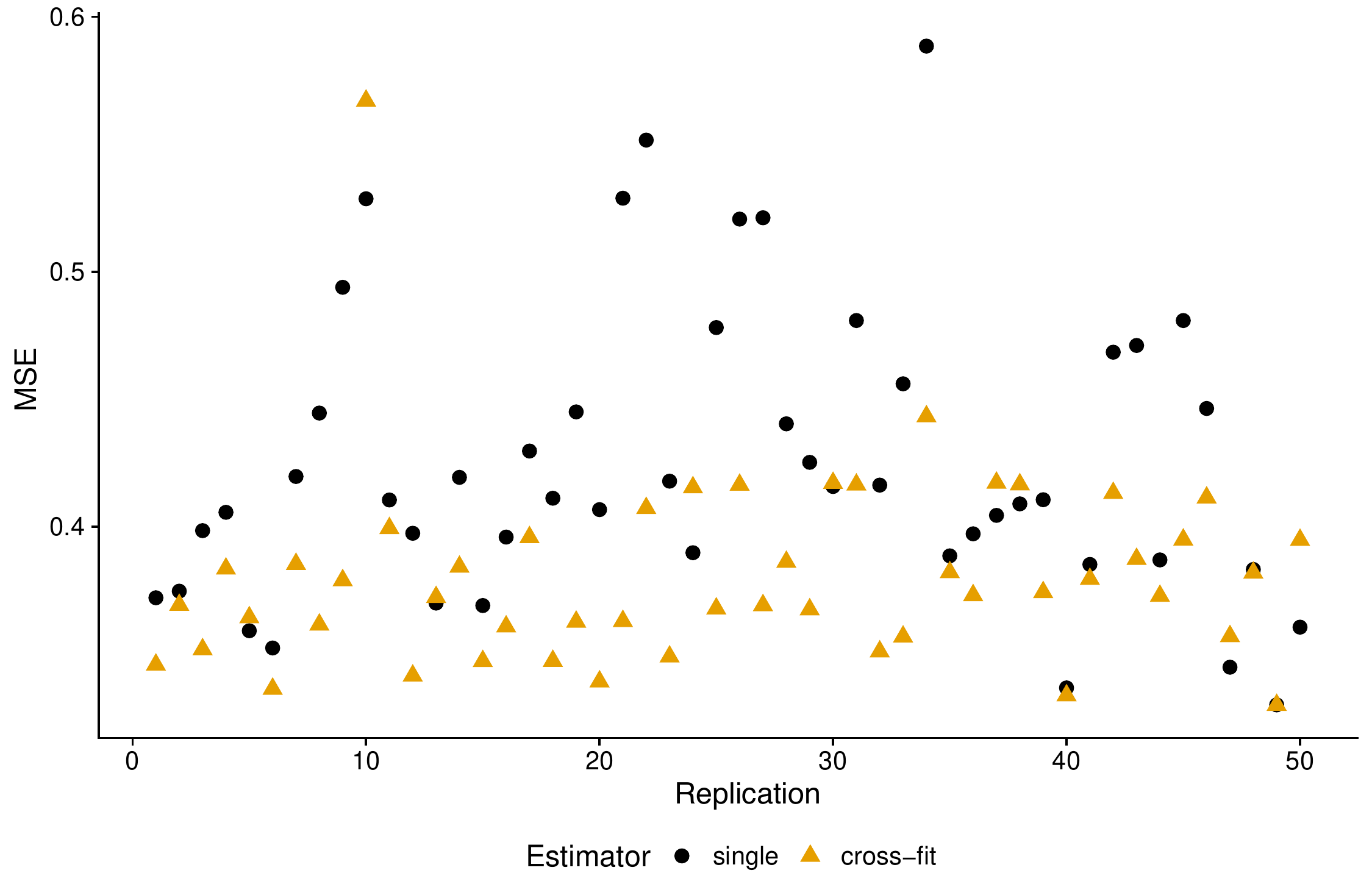}
\caption{Single vs. cross-fit estimation of CATE. \href{https://github.com/QuantLet/Meta_learner-for-Causal-ML/tree/main/Cross-Fitting}{\includegraphics[width=0.02\textwidth]{qletlogo_tr.png}$_{CF}$}}
\label{fig:single_vs_cross-fit}
\end{center}
\end{figure}

We give an example of the benefit from cross-fitting in Figure \ref{fig:single_vs_cross-fit}. We show the MSE from the true treatment effect for a single estimator and the cross-fit estimator based on a 50:50 sample split. We used the R-learner as the meta-learner and create 50 Monte Carlo replications of the data using the same data generating process (DGP) which simulates a RCT and has the following properties: $N = 2000$, $X = \mathbb{R}^{10}$, $e(X)$ = 0.5, and $\tau(x) = X_1 + \mathds{1}(X_2>0) + W \quad \text{with} \quad W \sim  \mathcal{N}(0,0.5)$. Using cross-fitting decreases the MSE compared to the single estimator in about 90\% of the cases. We also find that the variance is smaller compared to the single estimator.

In empirical studies we do not have an independent test set and setting aside a partition might not be efficient since we lose observations for the estimation part. In the following, we present an approach to use cross-fitting without an additional test set. 

\begin{figure}[ht]
\begin{center}
\includegraphics[width=0.9\textwidth]{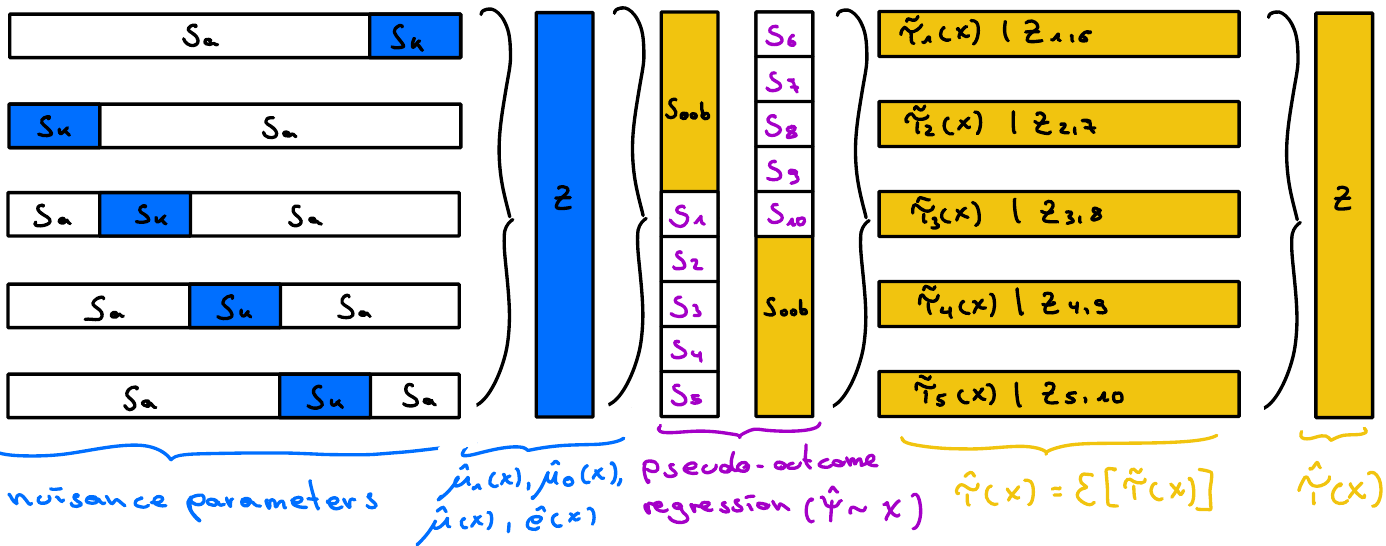}
\caption{Two-step sample splitting procedure.}
\label{fig:two-step-sample-split}
\end{center}
\end{figure}

We apply 5-fold sample splitting and use 80\% of the full data (denoted by $Z$) for training the nuisance functions (denoted by $S_a$) and 20\% to for estimation (denoted by $S_k$). We propose to estimate the nuisance parameters for each of the 5 folds, using all folds but $k$ for training and fold $k$ to predict the conditional mean and the propensity score. We then store the estimates. As a result, we have estimates of all nuisance parameters to create the pseudo-outcomes for each observation obtained from independent samples. Hence, the above-mentioned regularity conditions should be fulfilled. Now we want to train a regression model on the pseudo-outcomes (or minimize the R-loss). Instead of using the full sample for training and prediction, we divide the sample into different parts. We assign half of the sample as the test set (denoted by $S_{oob}$, which is short for out-of-bag) and the other half that is used to train the regression model. Let us say we want to rely on 5-fold cross-fitting (taking the average of $S_{oob}$ over 5 folds). We therefore split the other half of the sample into 5 folds (denoted by $S_{train} = \{S_1,S_2,\ldots,S_5\}$). Using each fold to train the regression model and predict on $S_{oob}$ leads to 5 estimates that we average by taking the mean. Now we reverse the role of $S_{train}$ and $S_{oob}$ and proceed as above. We apply this procedure to the DR- and R-learner. The S- and T-learner only needs one estimation step and hence it suffices to only use two different samples (the $S_a$ and $S_k$). In all other methods, we need an additional model (for example, the IPW-learner would also benefit from cross-fitting). The X-learner is quite robust even without cross-fitting. This might be because it only uses the propensity score in the last step. The advantage of the two-step sample splitting approach is that we have more observations to train the nuisance functions (in this example we have $S_a = 0.8Z$ observations instead of $0.8(Z-S_{oob})$). Figure \ref{fig:two-step-sample-split} shows the procedure in detail. As above, we denote $S_k$ as the fold which is used to estimate the nuisance parameters (e.g. the propensity score, the pseudo-outcomes). The estimators $\tilde{\tau}(x)$ refer to the CATE estimates obtained using $S_{oob}$ given different folds for training. For example,$\tilde{\tau}_1(x)$ first uses $S_1$ for training and $S_{oob}$ for estimation. To get estimates on the other half of the data (also denoted as $S_{oob}$) we use $S_6$ for training. Hence, $\tilde{\tau}_1(x)$ are estimates of the CATE for the whole dataset $Z$. A more detailed version of the cross-fitting part is shown in Figure \ref{fig:cross-fit_detailed} in the Appendix.

\vskip 0.5cm

There might be alternative ways and averaging procedures to ensure robust estimates and prediction results. For example, we could repeat the whole procedure $M$ times and generate different folds in the first place (the $S_k$ folds). The result would be $M$ estimates for each observation of $\hat{\tau}(x)$ over which we could take the median. This might lead to more robust estimates since it takes the sample splitting uncertainty into account. See \cite{jacob2020cross} for a Monte Carlo study about the implications of different sample-spitting, cross-fitting, and averaging approaches for meta-learner methods. The simulation study finds that the 5 fold cross-fitting with median averaging procedure works best. Our approach mimics this procedure but changes the way to define a test set on which the cross-fitting is applied. 
For the meta-learners, we have to do sample splitting and cross-fitting manually while the causal forest as well as the causal boosting relies on honest estimation and does sample splitting by default. Cross-fitting, as we define it, is not implemented in any of the modified ML methods.

\section{Empirical Examples}
To illustrate the methods presented in the previous sections, we consider two empirical examples. In the first example, we examine the effect that microcredits have on the total amount of loans, resulting from a randomized experiment in Morocco. In the second example, we study the effect of 401(k) eligibility on accumulated assets. This example deviates from random treatment assignment and contains self-selection into a treatment. While all presented methods condition on observed pre-treatment variables to estimate heterogeneity in treatment effects they should also be able to control for confounding variables. However, methods that use the propensity score should be more suited to eliminate the selection bias. For each method, we estimate the CATE and provide confidence intervals. We also show how to link the CATE to observed covariates for further analysis. In both examples, we apply the two-step sample splitting with a cross-fitting approach for the DR- and R-learner.

\subsection{Effect of microcredits on borrowing}
We start with an empirical dataset to analyze the effect of microcredit availability on borrowing activities such as the amount of loans (see \cite{crepon2015estimating} for a recent study using this dataset). Looking beyond the ATE and finding heterogeneous treatment effects is important to target specific groups and to make better policies. The allocation of treatment was randomized between 162 villages in Morocco. The villages were divided into pairs with similar observable characteristics. Then the treatment was randomly assigned to one of the pair while the counterpart was assigned to the control group. Treatment as microcredit availability in this context means that between 2006 and 2007 a microfinance institution started operating only in the treated villages. In 2009, 5,551 households were surveyed in a follow-up study. We use the results from this survey to estimate conditional average treatment effects using different methods and also show some strategies to get some insight into which characteristics are responsible for heterogeneity in treatment effects. 
We select the following pre-treatment covariates that are observed characteristics for each household such as the age of household's head, number of adults, number of children, total number of members in a household, indicators for households doing animal husbandry, other non-agricultural activity, household spouse responding to the survey, the education of the head and having an outstanding loan over the last year. 
Table \ref{tab:descriptive} shows the mean value for some covariates. They are categorized by all observations, the treated and the control group. Given these unconditional means, we see that the amount of loans for the treatment group is much higher ($2,930$) than for the control group ($1,802$). We also see that the mean of the characteristics is quite balanced across the two groups. This reassures us that the treatment assignment was randomly selected and that there are no confounding variables that lead to self-selection into treatment. While there are small differences in some covariates, this is not concerning since all methods that we apply make use of the propensity score or condition on the covariates to estimate the treatment effect only on similar subgroups. For example, more people in the treatment group already have a loan in the last twelve months. We can estimate the probability of being in the treatment group given this variable and reweigh the treatment and control group to adjust for these differences. The dataset and R-code for the microcredit analysis can be found here \href{https://github.com/QuantLet/Meta_learner-for-Causal-ML/tree/main/Microcredit-Example}{\includegraphics[height=5mm]{qletlogo_tr.png}$_{emp}$}. 

\begin{table}
\centering
\caption{Descriptive statistics of households (mean)}
\label{tab:descriptive}
\begin{tabular}{lrrr}
\hline \hline & All & Treated & Control \\
\hline \textit{Outcome Variable} & & & \\
Total Amount of Loans & 2,359 & 2,930 & 1,802 \\

\textit{Baseline Covariates} & & & \\
Number of Household Members & 3.879 & 3.872 & 3.886 \\
Number of Children & 1.266 & 1.261 & 1.272 \\
Head Age & 35.976 & 35.937 & 36.014 \\
Declared Animal Husbandry Self-employment Activity & 0.415 & 0.426 & 0.404 \\
Declared Non-agricultural Self-employment Activity & 0.146 & 0.129 & 0.164 \\
Borrowed from Any Source & 0.210 & 0.224 & 0.196 \\
Spouse of Head Responded to Self-employment Section & 0.067 & 0.074 & 0.061 \\
Member Responded to Self-employment Section & 0.044 & 0.048 & 0.041 \\

\hline \hline
\end{tabular}
\end{table}

We use three different ML algorithms to estimate the nuisance functions and to map the covariates on the pseudo-outcome (for the DR- and X-learner) or to minimize the R-loss function. Table \ref{tab:coef_micro} shows the coefficients for each ML algorithm obtained through cross-validation in the SuperLearner. The loss-function is the non-negative least squares based on the Lawson-Hanson algorithm which works for both Gaussian and binomial outcomes. We find that the neural network gets the highest weight for all functions except for the propensity score where the random forest has a slightly higher weight. Based on these weights, we only use the neural network and the random forest for the construction of the bootstrapped confidence intervals. This is mainly since we believe that even with a bootstrapped sample, the weights of the algorithms for each function will not change dramatically. Excluding the boosting algorithms decreases the computation time by about 50\%. 

\begin{table}[ht]
\centering
\caption{Weights of ML methods.}
\label{tab:coef_micro}
\begin{tabular}{rrrrrrr}
  \hline \hline
 & $\hat{e}(X)$ & $\hat{\mu}_0(X)$ & $\hat{\mu}_1(X)$ & $\hat{\mu}(X)$ & DR & R \\ 
  \hline
Boosting & 0.00 & 0.00 & 0.00 & 0.00 & 0.00 & 0.11 \\ 
Neural Network & 0.46 & 0.73 & 0.70 & 0.80 & 0.90 & 0.89 \\ 
Random Forest & 0.54 & 0.27 & 0.30 & 0.20 & 0.09 & 0.00 \\ 
   \hline \hline
\end{tabular}
\end{table}

\begin{table}[ht]
\centering
{\renewcommand{\arraystretch}{1.5}%
\begin{threeparttable}
\caption{CATE results for different methods.}
\label{tab:micro_CATE}
\begin{tabular}{llrrr}
\hline \hline
Category                                                                       & Method                                                                        & 20\% Least & ATE & 20\% Most \\ \hline
\multirow[t]{4}{*}{\textbf{Meta-Learner}}  & DR-learner         &  -15.4                   &1,119.8    & 3,057.6                  \\
            								& R-learner        	& 84.9                    &1,081.1     & 2,237.5                   \\
                							& T-learner         &  198.4                   & 1,152.5    &  2,470.3                  \\
                							& X-learner         &  869.9                  &1,137.2     &1,379.7   \\ \hline
\multirow[t]{2}{*}{\begin{tabular}[t]{@{}l@{}}\textbf{Modified ML}\\ \textbf{Methods}\end{tabular}} & Causal BART                                                                 & 593.8                    &1,132.3     &  2,304.7                  \\
                                                                               & Causal Forest &296.6                     &1,129.6     &2,329.6    \\
                                                                               \hline \hline               
\end{tabular}
\begin{tablenotes}
      \small
      \item \textit{} 
    \end{tablenotes}
  \end{threeparttable}
}
\end{table}

\begin{figure}[ht]
\begin{center}
\includegraphics[width=0.9\textwidth]{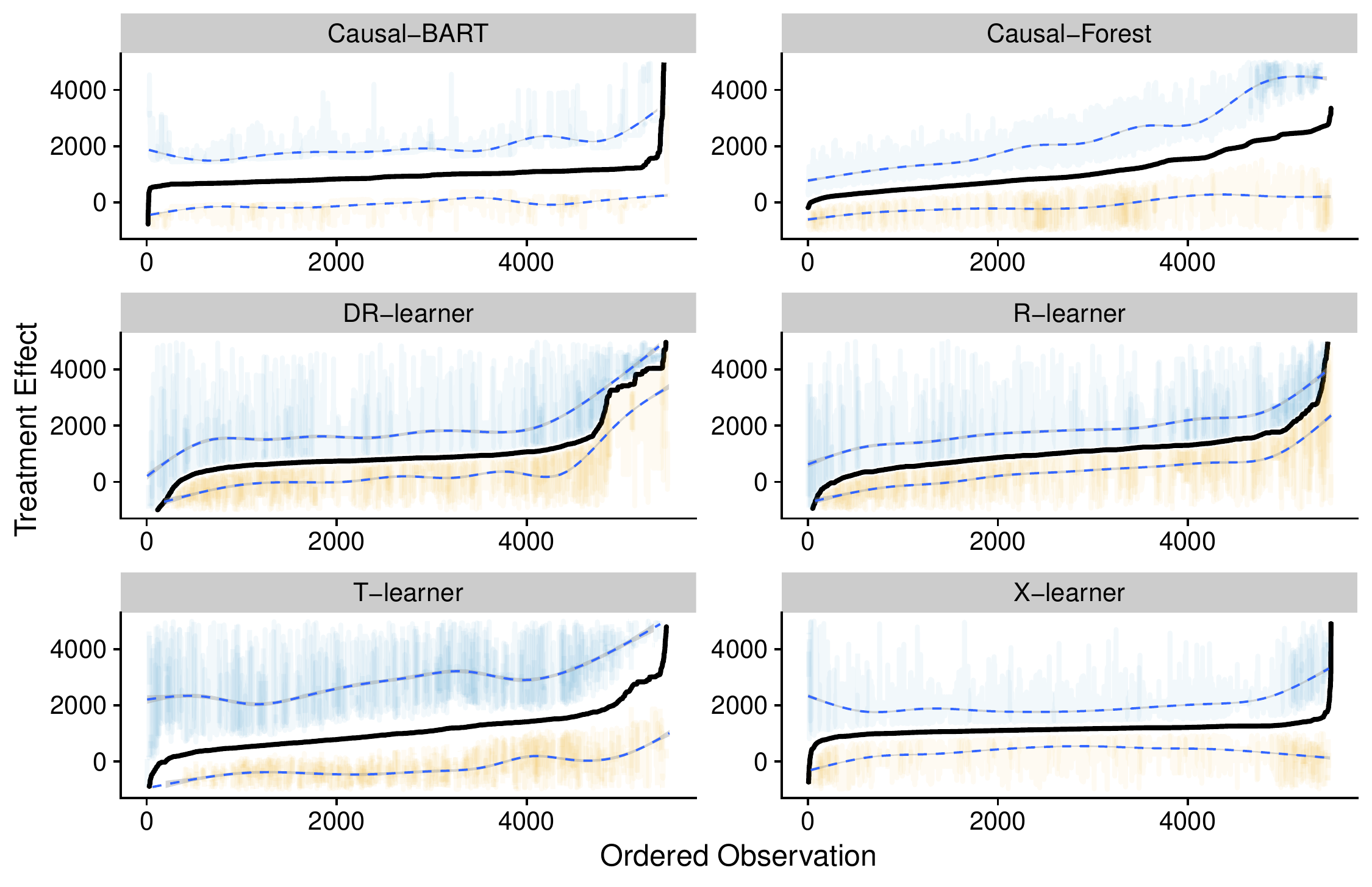}
\caption{Microcredit: Observations sorted by level of treatment effect.\href{https://github.com/QuantLet/Meta_learner-for-Causal-ML/tree/main/Microcredit-Example}{\includegraphics[width=0.02\textwidth]{qletlogo_tr.png}$_{sort}$}}
\label{fig:micro_CATE_sorted}
\end{center}
\end{figure}

Table \ref{tab:micro_CATE} shows a summary for the heterogeneous treatment, namely the effect for the 20\% least affected, the ATE, and the effect for the 20\% most affected observations. Especially for the quantiles, we find differences in the estimates given the methods that we consider. This holds for the lower 20\% where the effect ranges from $-15$ to $869$ as well as for the 20\% most affected with the lowest treatment effect from the X-learner with $1,379$ and the highest estimate from the DR-learner with $3,057$. The high value in the upper quantile from the DR-learner is because it predicts more extreme values at the tail of the distribution. The DR-learner also has the highest variance in terms of treatment effect with a range in number of loans from $- 15$ to $3,057$ on average for the specific quantiles. The ATE is around $1,100$ for all methods and there is no large difference between them. Figure \ref{fig:micro_CATE_sorted} shows the treatment effect for each observation, sorted by the size of the effect. We also show 95\% confidence intervals (CI). They are estimated via bootstrapping with $B$ = 500 replications. Here we adopt the procedure for the construction of CI's from \cite{kuenzel2019meta}. We first split our entire dataset into a training and validation set. We use the training set for bootstrapping by creating a sample from the training data of the same size with replacement. For each meta-learner and each bootstrap sample from the training data, we use the test set to estimate the CATE. We repeat this procedure $K =5$ times looping through all $k$ subsets and define them as the test set. In total, we end up having $B$ estimates for each observation on the whole data. Now we calculate the standard deviation ($\hat{\sigma}$) for each observation which we use to generate a lower and upper bound around the CATE estimates $[\hat{\tau}(x) - q_{\alpha/2}\hat{\sigma}; \hat{\tau}(x) + q_{1-\alpha/2}\hat{\sigma}]$. Especially for the meta-learners, we have a high variance between the bootstrapped samples indicating that even if the CATE is different, there might not be a significant heterogeneity. This is also in line with the estimates from the causal BART and the causal forest that show tighter bounds but also an almost flat CATE curve. To calculate the CATE, denoted by $\hat{\tau}(x)$, we use the whole training data, not a bootstrapped version, and proceed as described in the pseudo-code for the specific meta-learner.

Figure \ref{fig:micro_CATE_sorted} shows quite similar values for at least four of the methods. Only the DR- and R-learner have heavier tails for higher CATE estimates. The T-learner has the widest confidence intervals while all other methods show a similar range. Treatment effects based on the DR- and R-learner are heterogeneous, at least for the 20\% least and most affected. The most homogeneous prediction comes from the causal BART and the X-learner. Their estimates of the CATE are quite similar. The causal forest also shows an increasing slope of the point estimates but wide confidence intervals for the most affected observations. Based on these results, there is no clear evidence of treatment effect heterogeneity.  

In Figure \ref{fig:micro_CATE_sorted}, we sorted the treatment effect by its size for each method. This does not necessarily mean that all methods have the same order. To look into the order of the CATE based on each method, we show the correlation of the CATE among them in Figure \ref{fig:Microcredit_corrplot}. We show the Bravais-Pearson correlation coefficient ($\rho$), a histogram of the CATE, and correlation ellipses. It is reassuring that all methods are positively correlated. The highest correlation is between the doubly-robust and the R-learner ($\rho = 0.57)$ as well as between the T- and X-learner ($\rho = 0.5$). The smallest correlation appears between the causal BART and the R-learner with a correlation coefficient of $0.15$.  

If we believe that there is at least some difference in the effect between the least and most affected observations, then we can look at the average characteristics of these groups to understand what are potential drivers for the heterogeneity. Here we adopt a simple approach introduced by \cite{chernozhukov2018generic}, namely the Classification Analysis. The idea is to regress the least and most affected groups on some pre-chosen characteristics ($g(X)$) with $G_{q}$ being the observations given a specific group of the treatment effect:

\begin{center}
$\delta_{least} = \mathsf{E}[g(X) |G_{least}]$ \quad and  \quad $\delta_{most} = \mathsf{E}[g(X) |G_{most}]$.
\end{center}

\begin{table}[h]
\centering
{\renewcommand{\arraystretch}{1.5}%
\resizebox{0.9\textwidth}{!}{%
    \begin{threeparttable}
    \caption{Classification results for DR-learner and causal forest.}
\label{tab:micro_CLAN}
\begin{tabular}{lcccccc}
\hline \hline
                                                                                          & \multicolumn{3}{c}{DR-learner}                   & \multicolumn{3}{c}{Causal Forest}                 \\ \hline
                                                                                          & Most Affected & Least Affected & Difference      & Most Affected  & Least Affected & Difference      \\
\multirow[t]{3}{*}{Head age}                                                                 & 19.19 & 46.92 & -27.73          & 10.25 & 46.80 & -36.55  \\
                                                                                          & (17.81,20.58) & (45.53,48.30) & (9.271,11.22) & (45.82,47.77) & (-37.93,-35.17) & (-55.00,-51.49) \\
                                                                                          & -             & -              & {[}0.000{]}     & -              & -              & {[}0.000{]}     \\
\multirow[t]{3}{*}{\begin{tabular}[t]{@{}l@{}}Non-agricultural\\ self-employed\end{tabular}} & 0.118 & 0.186 & -0.068           & 0.073 & 0.136 & -0.064 \\
                                                                                          & (0.097,0.139) & (0.165,0.207) & (0.055,0.091) & (0.118,0.154) & (-0.089,-0.038)   & (0.121,0.232) \\
                                                                                          & -             & -              & {[}0.000{]}     & -              & -              & {[}0.000{]}     \\
\multirow[t]{3}{*}{\begin{tabular}[t]{@{}l@{}}Borrowed from \\ any source\end{tabular}}      & 0.138 & 0.338 & -0.201         & 0.050 & 0.388 & -0.338 \\
                                                                                             & (0.113,0.162) & (0.314,0.363) & (0.028,0.072) & (0.366,0.411) & (-0.370,-0.307) & (-0.351,-0.219) \\
                                                                                          & -             & -              & {[}0.000{]}     & -              & -              & {[}0.000{]}    \\ \hline \hline
\end{tabular}
\begin{tablenotes}
      \small
      \item \textit{Notes: 90\% confidence interval in parenthesis and p-values in brackets.} 
    \end{tablenotes}
  \end{threeparttable}
}
}
\end{table}

Here we focus on the head age, the probability of being self-employed in a non-agricultural sector, and whether someone had an outstanding loan over the past 12 months (borrowed from any source). In Table \ref{tab:micro_CLAN} we estimate the average of the characteristics for the two groups as well as if there is a significant difference between the groups. We show results for two methods, the doubly-robust meta-learner (DR-learner) and the causal forest. Detailed CLAN results that include all methods can be seen in the Appendix (\ref{tab:micro_CLAN_full}). For both methods, we find a significant difference in head age, probability of being self-employed in a non-agricultural sector, and probability of having a loan. The most affected people seem to be younger. Low values in the head age can arise since many people did not respond to that question and got a value of zero. However, we believe that the non-respondents are missing at random. This allows us to interpret the difference between the two. Looking at employment, we find diverse effects. Interpreting results from the DR-, X-learner, and causal forest we find that people who benefit most from microcredits are those who do not work in the non-agricultural sector. Results from the R-, T-learner and the causal BART suggest the other way around. We note that the result from the T-learner is not statistically significant. The absolute magnitude in the probability difference is rather small which is why we do not interpret this variable as a driver for treatment effect heterogeneity. We also find that people who already have a loan (with a higher probability) are less affected by microcredit availability. There are other possibilities to investigate which covariates might be drivers for effect heterogeneity. For example, if a tree-based method should be the best method for mapping the covariates on the treatment effects then we could use variable importance plots to see which variables (at a certain split in a tree) increase the variance between two leaves. If a variable is (randomly) chosen for a split and the mean values in the two resulting nodes are quite the same as before the split then this variable might not be very useful to explain the heterogeneity. We can also apply partial dependence plots to see how the treatment effect changes if we change one variable.

\subsection{Effect of 401(k) eligibility on accumulated assets}

While the microcredit data is based on a randomized controlled trial, the eligibility of a 401(k) pension plan is not. Only some firms offer access to a 401(k) and hence there is self-selection into treatment. It might be the case that more educated people chose firms that provide a 401(k) pension plan and that they have higher financial assets in the first place. \cite{poterba1994401} argue that conditioning on observed characteristics, like the income, can restore the random assignment mechanism. The dataset we use is the same as in \cite{chernozhukov2004effects} which is based on the 1991 Survey of Income and Program Participation. We are interested in the question if 401(k) eligibility, our treatment variable, has an impact on accumulated assets (here we use the net financial assets as the outcome variable). We control for income and other variables related to the job choice that may have an impact on treatment assignment and assets. In total, we have 9,915 observations and 13 covariates consisting of age, family size, income, years of education, and indicator variables for married, two-earner status, defined benefit pension status, homeownership, and IRA participation. We split the dataset into 5 parts and proceed as described in the sample splitting section (\ref{sec:sample_splitting}). The dataset and R-code for the 401(k) analysis can be found here \href{https://github.com/QuantLet/Meta_learner-for-Causal-ML/tree/main/401k-Example}{\includegraphics[height=5mm]{qletlogo_tr.png}$_{emp}$}.

\begin{table}[ht]
\centering
\caption{Descriptive statistics of observations (mean)}
\label{tab:401k_descriptive}
\begin{tabular}{lrrr}
\hline \hline & All & Treated & Control \\
\hline \textit{Outcome Variable} & & & \\
Net financial assets & 18,052 & 30,347 & 10,788 \\

\textit{Baseline Covariates} & & & \\
Age & 41.06 & 41.48 & 40.81 \\
Income & 37,201 & 46,862 & 31,494 \\
Years of education & 13.21 & 13.76 & 12.88 \\
Proportion of being married & 0.60 & 0.67 & 0.56 \\
Proportion of two-earners & 0.38 & 0.48 & 0.31 \\
Proportion of home-ownership & 0.63 & 0.74 & 0.57 \\
\hline \hline
\end{tabular}
\end{table}

Table \ref{tab:401k_descriptive} shows the mean values for the net financial assets and for some pre-treatment covariates. The amount of assets is higher in the treatment group than in the control group. Concerning the self-selection into treatment, we see that some characteristics are different between the treatment and control group. For example, the proportion of home-ownership, years of education, and income is higher for treated people. There are further reasons to believe that such characteristics are positively correlated with financial assets. In this case, we have to control for such variables to account for the self-selection into treatment.

\begin{table}[]
\centering
\caption{Weights of ML methods.}
\label{tab:coef_401}
\begin{tabular}{rrrrrrr}
  \hline \hline
 & $\hat{e}(X)$ & $\hat{\mu}_0(X)$ & $\hat{\mu}_1(X)$ & $\hat{\mu}(X)$ & DR & R \\ 
  \hline
Boosting & 0.36 & 0.08 & 0.05 & 0.08 & 0.00 & 0.00 \\ 
Neural Network & 0.14 & 0.09 & 0.06 & 0.06 & 0.49 & 0.33 \\ 
Random Forest & 0.50 & 0.82 & 0.89 & 0.85 & 0.51 & 0.67 \\  
   \hline \hline
\end{tabular}
\end{table}

\begin{table}[ht]
\centering
{\renewcommand{\arraystretch}{1.5}%
\begin{threeparttable}
\caption{CATE results for different methods.}
\label{tab:401k_CATE}
\begin{tabular}{llrrr}
\hline \hline
Category                                                                       & Method                                                                        & 20\% Least & ATE & 20\% Most \\ \hline
\multirow[t]{4}{*}{\textbf{Meta-Learner}}  & DR-learner         &  4,998                   &7,120     & 9,806                   \\
            								& R-learner        	& 4,250                    &7,410     & 11,320                   \\
                							& T-learner         &  -4,171                  & 7,579    &  25,326                  \\
                							& X-learner         &  -285                 &7,631     &18,648   \\ \hline
\multirow[t]{2}{*}{\begin{tabular}[t]{@{}l@{}}\textbf{Modified ML}\\ \textbf{Methods}\end{tabular}} & Causal BART                                                                 & 2,466                    &9,055     &  21,525                  \\
                                                                               & Causal Forest &5,210                     &8,228     &12,360    \\
                                                                               \hline \hline               
\end{tabular}
\begin{tablenotes}
      \small
      \item \textit{} 
    \end{tablenotes}
  \end{threeparttable}
}
\end{table}

\begin{figure}[ht!]
\begin{center}
\includegraphics[width=0.9\textwidth]{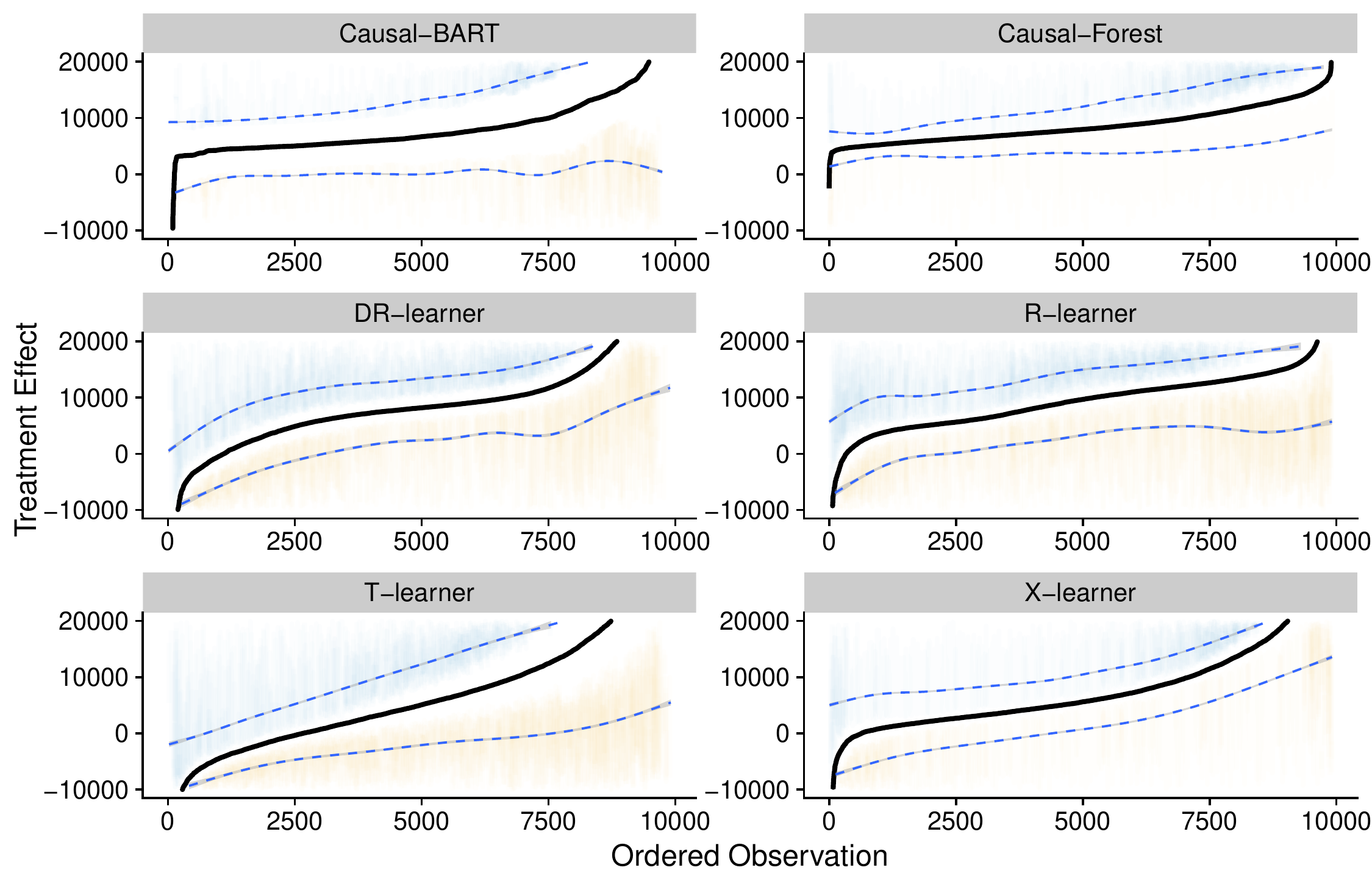}
\caption{401k: Observations sorted by level of treatment effect.\href{https://github.com/QuantLet/Meta_learner-for-Causal-ML/tree/main/401k-Example}{\includegraphics[width=0.02\textwidth]{qletlogo_tr.png}$_{sort}$}}
\label{fig:401k_CATE_sorted}
\end{center}
\end{figure}

Table \ref{tab:401k_CATE} shows the estimated CATE for the 20\% least affected and 80\% most affected as well as the ATE. The ATE is positive and ranges from 7,120 to 9,055, depending on the method. Its variance between the methods is quite low, compared to the estimates for the least and most affected groups. While the T- and X-learner predict a negative effect from 401(k) eligibility on financial assets for the lowest group, all other methods predict a positive effect. The highest affected group has values from 9,806 (from the DR-learner) to 25,326 (from the T-learner). The causal forest predicts values with the lowest heterogeneity. Except for the causal forest, all other learners predict extreme values in the tails of the CATE. If we would use a majority vote from all the methods to interpret the estimated effects, then it is reassuring that everyone has a positive effect from the 401(k) eligibility as can be seen in Figure \ref{fig:401k_CATE_sorted}. Given the wide confidence intervals, the evidence of treatment effect heterogeneity is not so clear. 

Figure \ref{fig:401k_corrplot} shows the correlation of the CATE between the different methods. We find that the methods are highly correlated with each other. The lowest correlation is between the DR- and T-learner with a correlation coefficient of $\rho= 0.64$ while the highest correlation is between the causal BART and causal forest ($\rho=0.85$). The reason why the estimated CATE is more similar might be the large sample size of $N=9915$. 

\begin{figure}[ht!]
\centering
\includegraphics[width=0.9\textwidth]{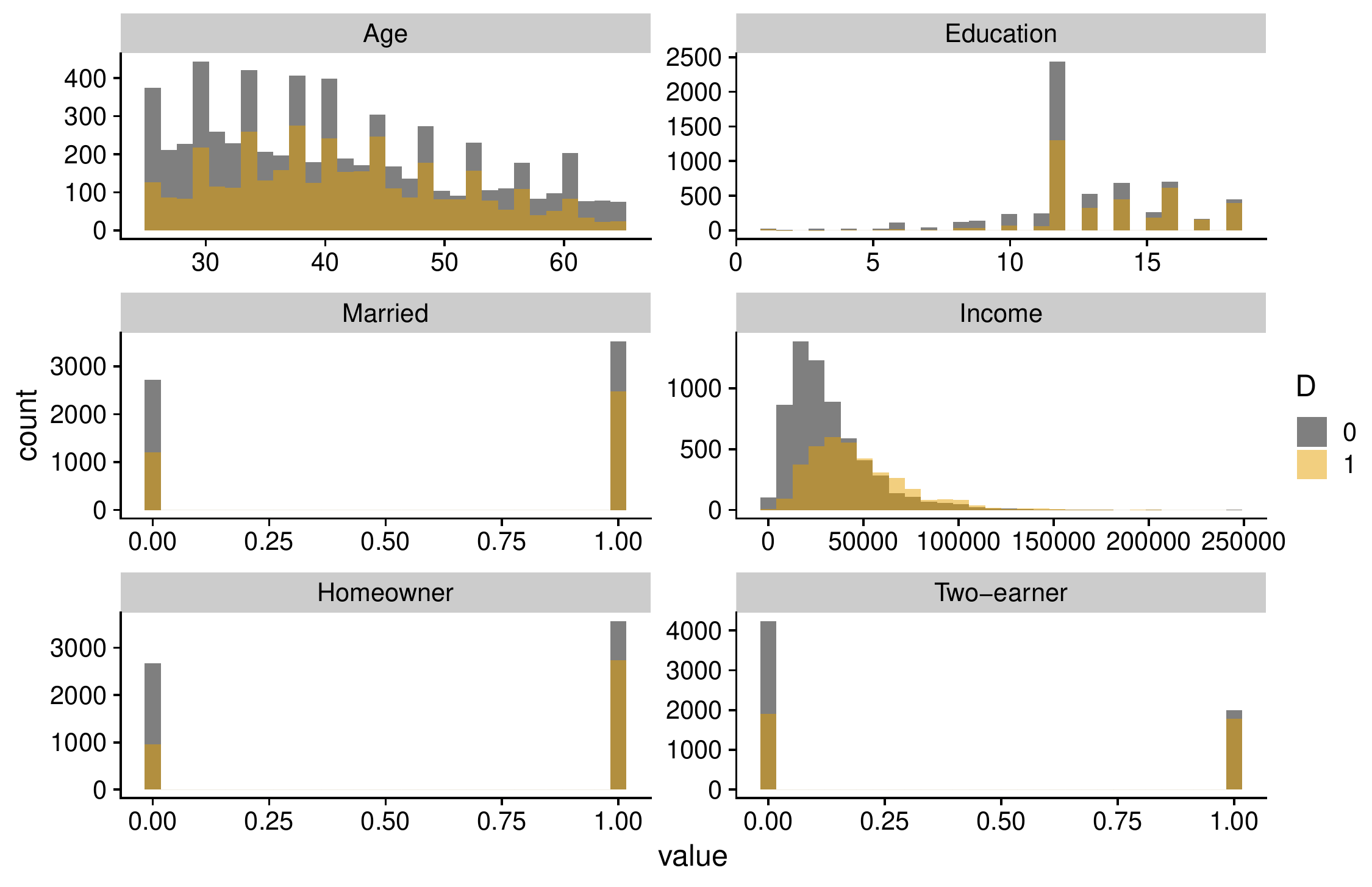}
\caption{Unweighted distribution of variables given treatment status.\href{https://github.com/QuantLet/Meta_learner-for-Causal-ML/tree/main/401k-Example}{\includegraphics[width=0.02\textwidth]{qletlogo_tr.png}$_{nb}$}}
\label{fig:401k_balance_noweight}
\end{figure}

\begin{figure}[ht!]
\centering
\includegraphics[width=0.9\textwidth]{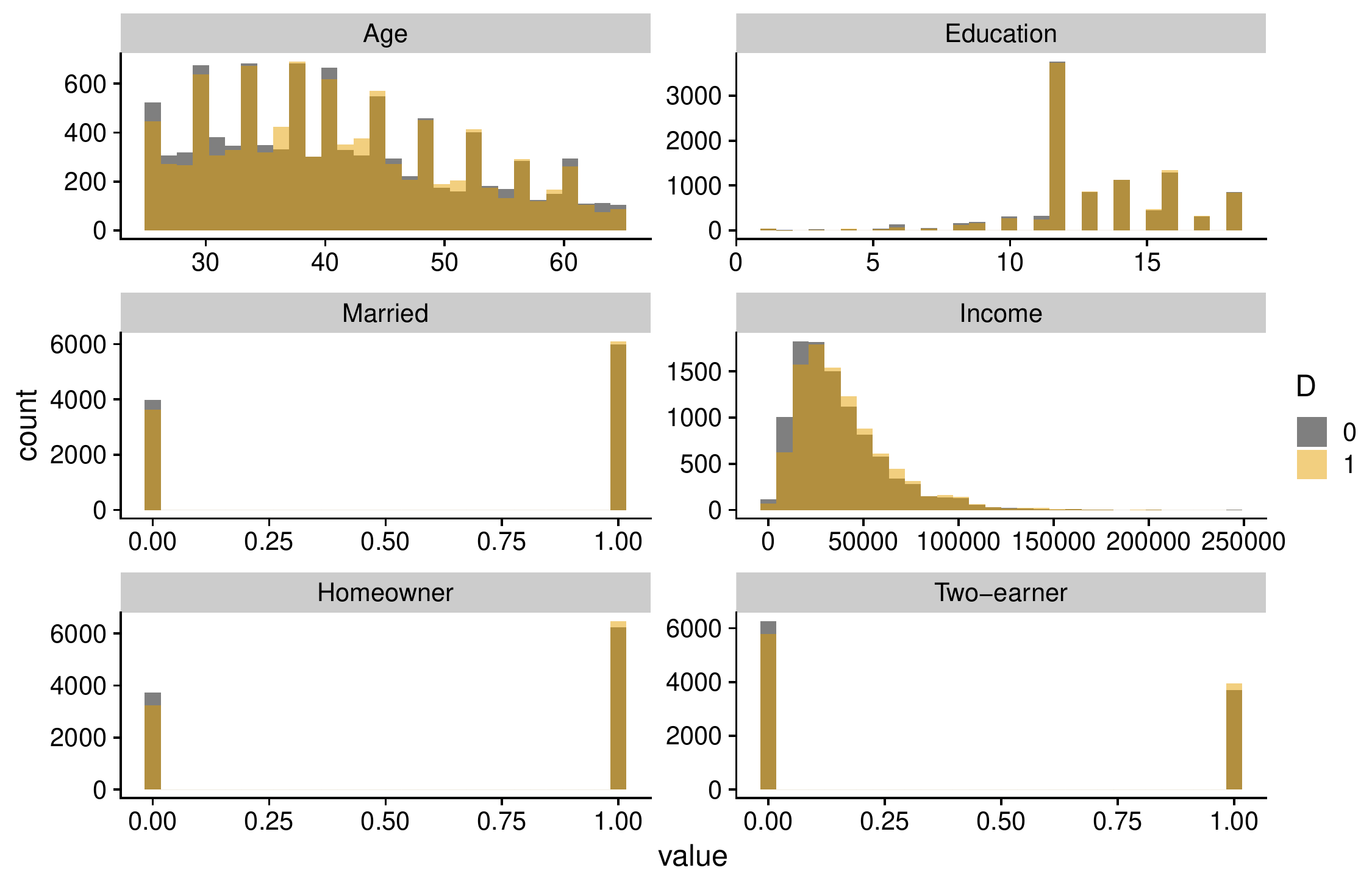}
\caption{IPW weighted distribution of variables given treatment status.\href{https://github.com/QuantLet/Meta_learner-for-Causal-ML/tree/main/401k-Example}{\includegraphics[width=0.02\textwidth]{qletlogo_tr.png}$_{IPW}$}}
\label{fig:401k_balance_IPW}
\end{figure}

Since the data does not come from a randomized controlled trial, we expect the distribution of the covariates to be different given treatment status. To see this, we plot the distribution of age, years of education, marital status, income, homeowner status, and two-earner status in Figure \ref{fig:401k_balance_noweight}. As we already saw from Table \ref{tab:401k_descriptive}, treated people have a higher income, slightly more years of education and, among others, are more often homeowners. To see if the estimated propensity score can catch the differences, we can look at a weighted histogram. What we do is weigh the counts in each variable by the inverse of the propensity score. If someone is in the treatment group, we weigh by $1/\hat{e}(x)$ and the control group observations by  $1/(1-\hat{e}(x))$. The result is shown in Figure \ref{fig:401k_balance_IPW}. Indeed, we see that the distributions are quite similar after reweighing with the propensity score.

\section{Simulated Data}

Since the true treatment effect is never known beforehand, we provide a simulation to evaluate different approaches in terms of performance for parameter estimation. The data-generating process allows controlling the number of observations, dimensionality, and the distributions of the variables. The possibility to specify datasets for different simulations and scenarios helps to investigate the methods used in this tutorial. Note that simulated data often lack realistic data structures. An alternative is to rely on synthetic data where only the treatment effect is artificially added. Since a simulation study in the type of a Monte Carlo study is not the main focus of this tutorial, we will only use two simulated data generating processes. The purpose is to give an idea of how to simulate data and test different methods. Instead of relying on purely artificial data, \cite{wendling2018comparing} creates synthetic data based on real covariates and a treatment assignment mechanism. Only the outcome is simulated based on non-parametric models of the real outcome.  

\subsection{Data Generating Process}
 
The basic model used in this tutorial is a partially linear regression model based on \cite{robinson1988root} with extensions:  

\begin{align}
Y = \tau(X_i)&D + \mu_0(X_i) + U,  &&\mathsf{E}[U | X,D] = 0, \\
&D = e(X_i) + V,  &&\mathsf{E}[V | X] = 0, \label{prop_score} \\
\tau(X_i) &= t(X_i) + W, &&\mathsf{E}[W| X] = 0.
\end{align}

Let $Y$ be the outcome variable. $\tau(X_i)$ is the true treatment effect or population uplift, while $D$ is the treatment status. The vector $X = ({X_{1},...,X_{p}})$ consists of $p$ different features or covariates and $U$, $V$ and $W$ are unobserved covariates which follow a random normal distribution = $N(0,1)$.  

Equation \ref{prop_score} is the propensity score. In the case of completely random treatment assignment, the propensity score is constant for all units, and, if equally distributed, then ${e}(X_{i}) = 0.5$. The covariates $X$ are generated from a random multivariate normal distribution ($N(0,1)$). Note that all values are continuous. In business applications, discrete values (categorical variables) are very common. For the data generation process as well as for the evaluation of most models, it would make no difference if such variables are present. This is because vanilla machine learning methods can handle categorical variables quite well. An exception is the causal forest where one has to use one-hot encoding, to transform the variable into dummies. Next, we describe the generation of the functions in detail. \\

\begin{tcolorbox}[sharp corners]
\begin{center}
\underline{Covariates (X)}
\end{center}
\begin{enumerate}
\item Generate a random positive definite covariance matrix $\Sigma$ based on a uniform distribution over the space $p \times p$ of the correlation matrix. Let $p = 20$.  
\item Scale the covariance matrix. This equals the correlation matrix and can be seen as the covariance matrix of the standardized random variables $\Sigma = \frac{X}{\sigma(X_i)}$. 
\item Generate random normal distributed variables $X_{N \times p}$ with mean = 0 and variance = $\Sigma$.
\end{enumerate}
\end{tcolorbox}

The function $\mu_{0}(X_i)$ is calculated using a linear function with interaction terms and contains the following covariates:

\begin{align}
\mu_{0}(X) &= X_1 \otimes X_{2} + X_{3} \otimes X_4 + X_5.
\end{align}

All covariates are normal distributed except $X_5$ which only takes four values, namely $\{-0.2,0,0.2,0.6\}$. 
Next, we describe how to build the function $e(X_i)$ as well as how to create heterogeneous treatment effects. A varying treatment effect implies that its strength differs among the observations and is therefore conditioned on some covariates $X$. 
Regarding the treatment assignment, two settings are considered. Setting 1 assumes $D$ to be completely randomly assigned among the observations. In this case, $D$ is just a vector of random numbers with values $0$ or $1$.  In setting 2, the treatment assignment is dependent on the covariates. The binary treatment assignment is generated through a Bernoulli function. This implies per default a sort of uncertainty or random error. Even if the probability from the propensity score is 90\% for $D = 1$, there is still a 10\% chance that it is generated to be zero. The functions are generated as follows: \\

\begin{tcolorbox}[sharp corners]
\begin{center}
\underline{Treatment Assignment ($e_0$)}
\end{center}
\underline{Setting 1: $e_0$}
\begin{align*}
D  \overset{ind.}{\sim} \text{Bernoulli}(e_0), &&\text{with} \quad e_0 = 0.5
\end{align*}
\underline{Setting 2: $e(X_i)$}
\begin{enumerate}
\item Dependence is non-linear (through interaction terms): $a(X_i) = X_1 \otimes X_{2} + X_{3} \otimes X_{4}$.
\item Calculate the probability distribution for the vector $a$ from the normal distribution function:  
\begin{align*}
e(X_i) &= \Phi\left(\frac{a-\mu(a)}{\sigma(a)} \right) = \frac{1}{2} \left[1 + \operatorname{erf}\left(\frac{a-\mu(a)}{\sigma(a) \sqrt 2 }\right)\right] 
\end{align*}
Here $\mu(a)$  denotes the mean of $a$ and $\sigma(a)$ the standard deviation. 
\item Apply a random number generator from a Binomial function $B\{N,e(X_i)\}$ with probability for success = $e(X_i)$. This creates a vector $D \in \{0;1\}$ such that $D \overset{ind.}{\sim} \text{Bernoulli}\{e(X_i)\}$.
\end{enumerate}
\end{tcolorbox}

Regarding the treatment effect, we also consider two different settings. First, $\tau(X_i)$ depends linear on covariates $X$, and second, $\tau(X_i)$ has a non-linear, more complex form concerning the covariates. In both settings, we can examine heterogeneous treatment effects. The vector $b = \frac{1}{l}$ with $l \in \{1,2,...,p\}$ represents weights for every covariate.  \\

\begin{tcolorbox}[sharp corners]
\begin{center}
\underline{Treatment Effect ($\tau(X_i)$)}
\end{center}
\underline{Setting 1: linear dependence}\\
\begin{center}
$\tau(X_i) = 0.6X_1 + 0.6X_2 + 0.6X_3 + 0.6X_4 + X_5 + W \quad \text{with} \quad W \widesim{}  \mathcal{N}	(0,0.5)$. \\

\end{center}
\underline{Setting 2: non-linear dependence} \\
\begin{center}
$\tau(X_i) = \sin(X_{1:3} \times b_{1:3}) +  1.5\cos( X_{4}) + X_5$.

\end{center}

\end{tcolorbox}

The simulated data that include the true treatment effect can be found here: \href{https://github.com/QuantLet/Meta_learner-for-Causal-ML/tree/main/Simulation-Example}{\includegraphics[height=5mm]{qletlogo_tr.png}$_{sim}$}.

\subsection{Results}

To evaluate the different methods, we consider two data generating processes (DGP). Setting 1 is a randomized controlled trial with a constant propensity score of 0.5, while the treatment effects depend linear on covariates. In setting 2, we consider confounding variables, namely that the treatment probability now depends on covariates (through interactions of covariates) while the treatment effect depends non-linear on the covariates. In both settings, we set $N = 2000$ and $p = 20$. We use up to 5 variables to generate the different variables and the treatment effects while all other variables have no dependence on any function. They are spurious and the hope is that the ML methods find the important variables while excluding the others. Table \ref{tab:sim_CATE} shows the mean squared error (MSE) for all considered methods and both settings. We list to different versions of the DR- and R-learner. The first is the in-sample estimator where the regression and estimation of the CATE based on the pseudo-outcome or R-loss is done on the same sample. Only the nuisance functions are regressed and estimated on different samples. This is in line with the sample splitting theory. In the last step, we just want to have a good approximation of the CATE which is why we can use the same sample for training and prediction. Note that the DR-learner already estimates the CATE in the pseudo-outcome. Using the whole sample should increase the prediction power. 

In practice, however, we find that it might be better to split the sample again and not use the same sample for training and prediction. The reason is the following: If the predictions of the nuisance functions are not perfect, the pseudo-outcome deviates from the true CATE. The deviation becomes clearer with a higher estimation error and also if there are extreme values in the propensity score. Using different samples in the last step aims to smooth the function and discard outliers. This approach adds sample splitting (the two-step sample splitting without cross-fitting). Here we apply this approach with cross-fitting. This means we not only want to have different samples for training and prediction but also want to average the prediction fold over different training samples. Therefore we split the sample into  6 folds (the proportions are 10\% for the first 5 folds and 50\% for the last one). We use fold 1 to 5 individually to train regression functions and predict on all fold 6. Then we average the 5 estimates for fold 6. Now we reverse the role, combining fold 1 to 5 and split fold 6 in 5 parts. We proceed as above. We call the sample split estimator simply cross-fit estimator. Table \ref{tab:sim_CATE} shows the results for both versions. Using the cross-fit version we can decrease the MSE by at least 50\%. In simulations, we find that even a 50:50 split where we use 50\% for training and predict on the other half can decrease the MSE. The cross-fit version turns out to further decrease the MSE in simulations with different DGP's. For completeness, we show the procedure of the 50:50 split approach in Figure \ref{fig:two-step_50_50} in the Appendix. 

Table \ref{tab:coef_sim} shows that in setting 1 the tree-based methods (boosting and random forest) perform best in predicting the propensity score while the neural network does better in the regression tasks. The lasso only gets significant weight in the treatment effect regression. The lasso is excluded when applying the R-learner since in a linear setting the loss-function slightly differs from the more general non-parametric one. If the data generating process becomes more complex, the lasso method becomes less important shifting weights towards the neural network. In setting 2, the tree-based methods are most important in all tasks but for the treatment-effect regression based on the R-learner. We also experimented with excluding the neural network and found that in the linear setting, more weight is based on the lasso, while in the non-linear setting, the tree-based methods are superior. Since all methods are important in at least one task, we include all methods but the lasso when creating bootstrapped confidence intervals. 

\begin{table}[ht]
\centering
\caption{Weights of ML methods.}
\label{tab:coef_sim}
\begin{tabular}{rrrrrrr}
  \hline \hline
 & $\hat{e}(X)$ & $\hat{\mu}_0(X)$ & $\hat{\mu}_1(X)$ & $\hat{\mu}(X)$ & DR & R \\ 
  \hline
  \textit{Setting 1} &&&&&& \\
Boosting 		& 0.91 & 0.11 & 0.27 & 0.27 & 0.26 & 0.12 \\ 
Lasso 			& 0.00 & 0.00 & 0.01 & 0.00 & 0.34 & --- \\
Neural Network 	& 0.00 & 0.83 & 0.70 & 0.68 & 0.00 & 0.75 \\ 
Random Forest 	& 0.09 & 0.06 & 0.02 & 0.04 & 0.39 & 0.12 \\ 
\hline
  \textit{Setting 2} &&&&&& \\
Boosting & 0.40 & 0.62 & 0.72 & 0.72 & 0.14 & 0.13 \\ 
Lasso 			& 0.00 & 0.00 & 0.00 & 0.00 & 0.09 & --- \\
Neural Network & 0.00 & 0.07 & 0.06 & 0.07 & 0.13 & 0.39 \\ 
Random Forest & 0.60 & 0.31 & 0.22 & 0.21 & 0.65 & 0.47 \\ 
  
   \hline \hline
\end{tabular}
\end{table}

\begin{table}[ht]
\centering
{\renewcommand{\arraystretch}{1.5}%
\resizebox{0.8\textwidth}{!}{%
    \begin{threeparttable}
\caption{MSE for different methods. \href{https://github.com/QuantLet/Meta_learner-for-Causal-ML/tree/main/Simulation-Example}{\includegraphics[width=0.02\textwidth]{qletlogo_tr.png}$_{res}$}}
\label{tab:sim_CATE}
\begin{tabular}{llrr}
\hline \hline
Category                                                                       & Method                                                                        & MSE Setting 1& MSE Setting 2 \\ \hline
\multirow[t]{4}{*}{\textbf{Meta-Learner}}  	& DR-learner (in-sample)   	&2.68     	&  3.42                                        \\
											& DR-learner (cross-fit)  	&1.11     	&  1.17 \\
            								& R-learner (in-sample)   	&2.36   	&  2.58                                         \\
            								& R-learner (cross-fit)  	&0.80      	&  1.34  \\
                							& T-learner    				&1.17     	&  2.09                                        \\
                							& X-learner    				&0.58    	&  1.10                  \\ \hline
\multirow[t]{2}{*}{\begin{tabular}[t]{@{}l@{}}\textbf{Modified ML}\\ \textbf{Methods}\end{tabular}} & Causal BART   &0.55  & 0.48                                    \\
                                                                               & Causal Forest &0.86  &1.34                          \\
                                                                               \hline \hline               
\end{tabular}
\begin{tablenotes}
      \small
      \item \textit{Notes: The term oob denotes that the sample is split while in-sample denotes the use of the whole sample.} 
    \end{tablenotes}
  \end{threeparttable}
}
}
\end{table}

\begin{figure}[ht]
\begin{center}
\includegraphics[width=0.8\textwidth]{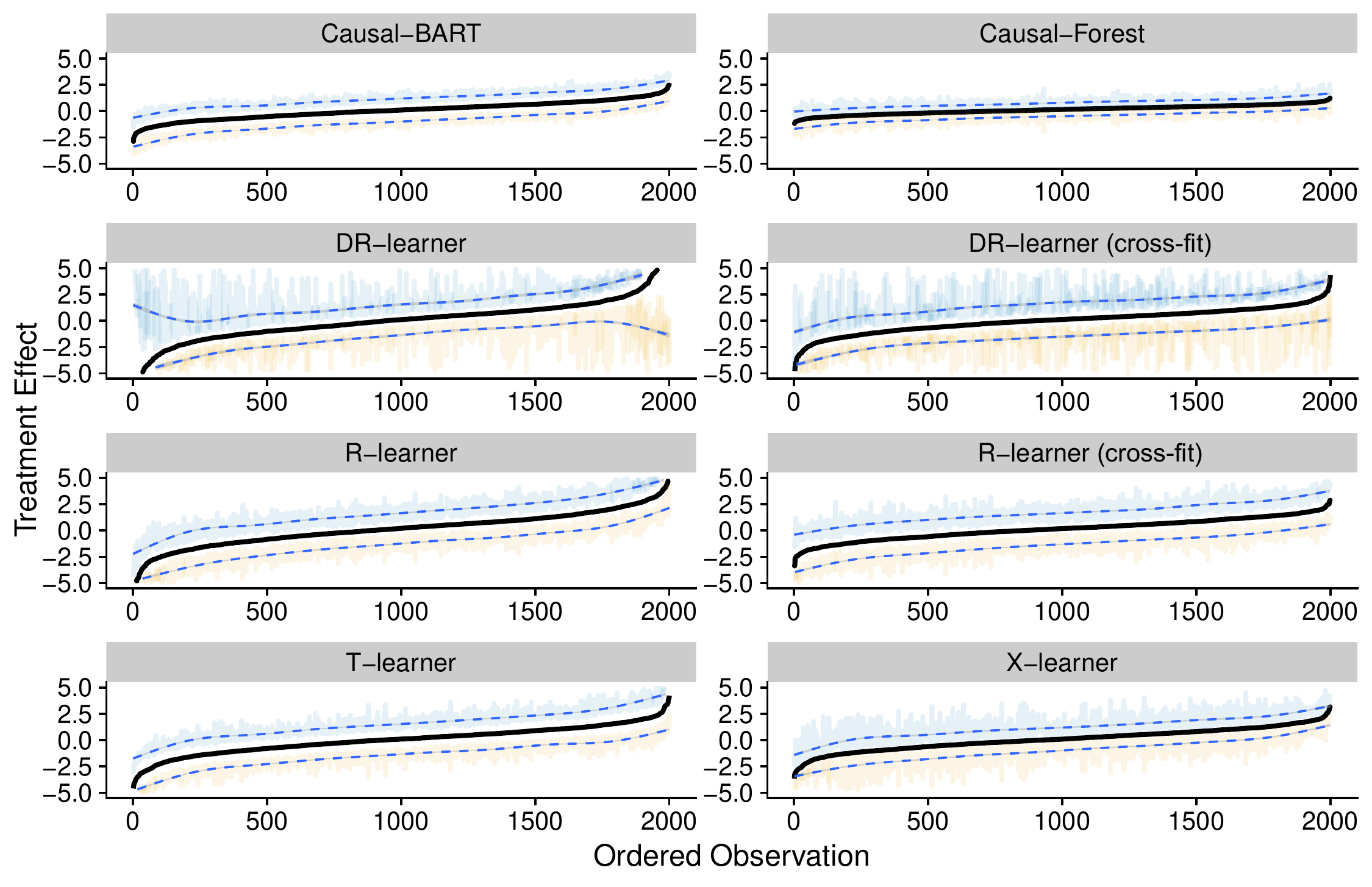}
\caption{Setting 1: Observations sorted by treatment effect. \href{https://github.com/QuantLet/Meta_learner-for-Causal-ML/tree/main/Simulation-Example}{\includegraphics[width=0.02\textwidth]{qletlogo_tr.png}$_{S1}$}}
\label{fig:sim_CATE_sorted_S1}
\end{center}
\end{figure}

\begin{figure}[ht]
\begin{center}
\includegraphics[width=0.8\textwidth]{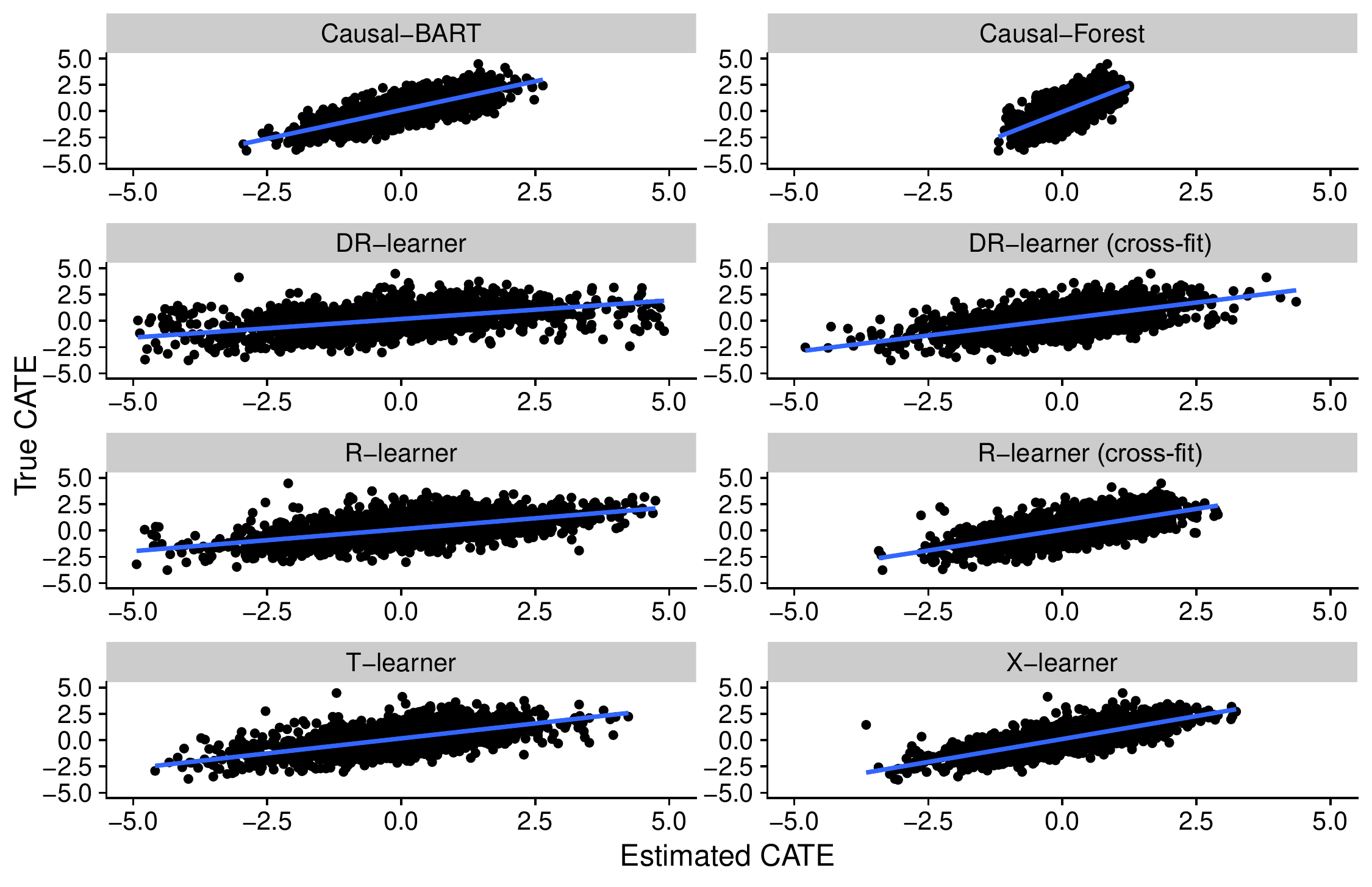}
\caption{Setting 1: Scatterplot of estimated and true CATE. \href{https://github.com/QuantLet/Meta_learner-for-Causal-ML/tree/main/Simulation-Example}{\includegraphics[width=0.02\textwidth]{qletlogo_tr.png}$_{T1}$}}
\label{fig:sim_CATE_scatter_S1}
\end{center}
\end{figure}

\begin{figure}[ht]
\begin{center}
\includegraphics[width=0.8\textwidth]{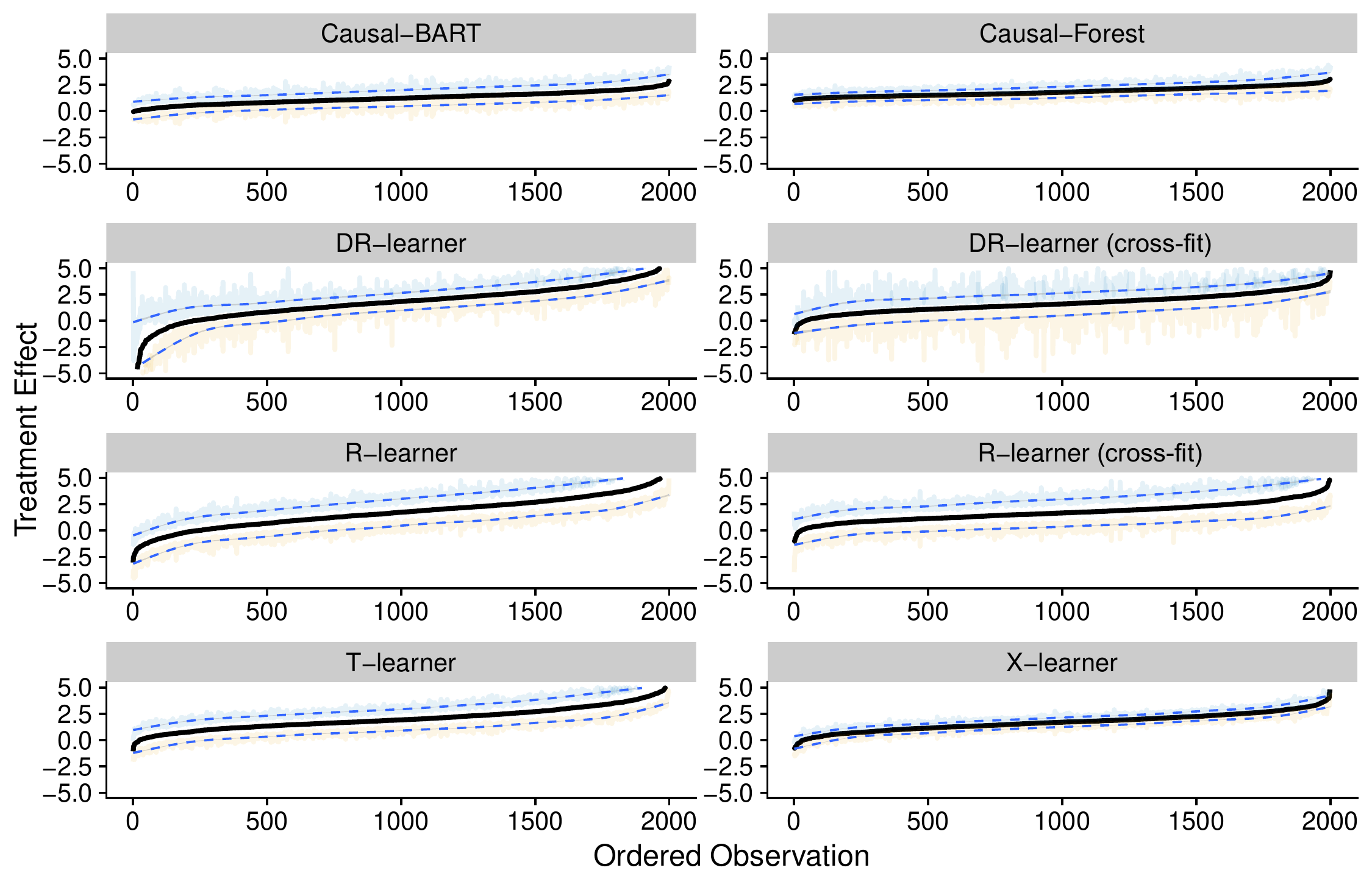}
\caption{Setting 2: Observations sorted by treatment effect. \href{https://github.com/QuantLet/Meta_learner-for-Causal-ML/tree/main/Simulation-Example}{\includegraphics[width=0.02\textwidth]{qletlogo_tr.png}$_{S2}$}}
\label{fig:sim_CATE_sorted_S2}
\end{center}
\end{figure}

\begin{figure}[ht]
\begin{center}
\includegraphics[width=0.8\textwidth]{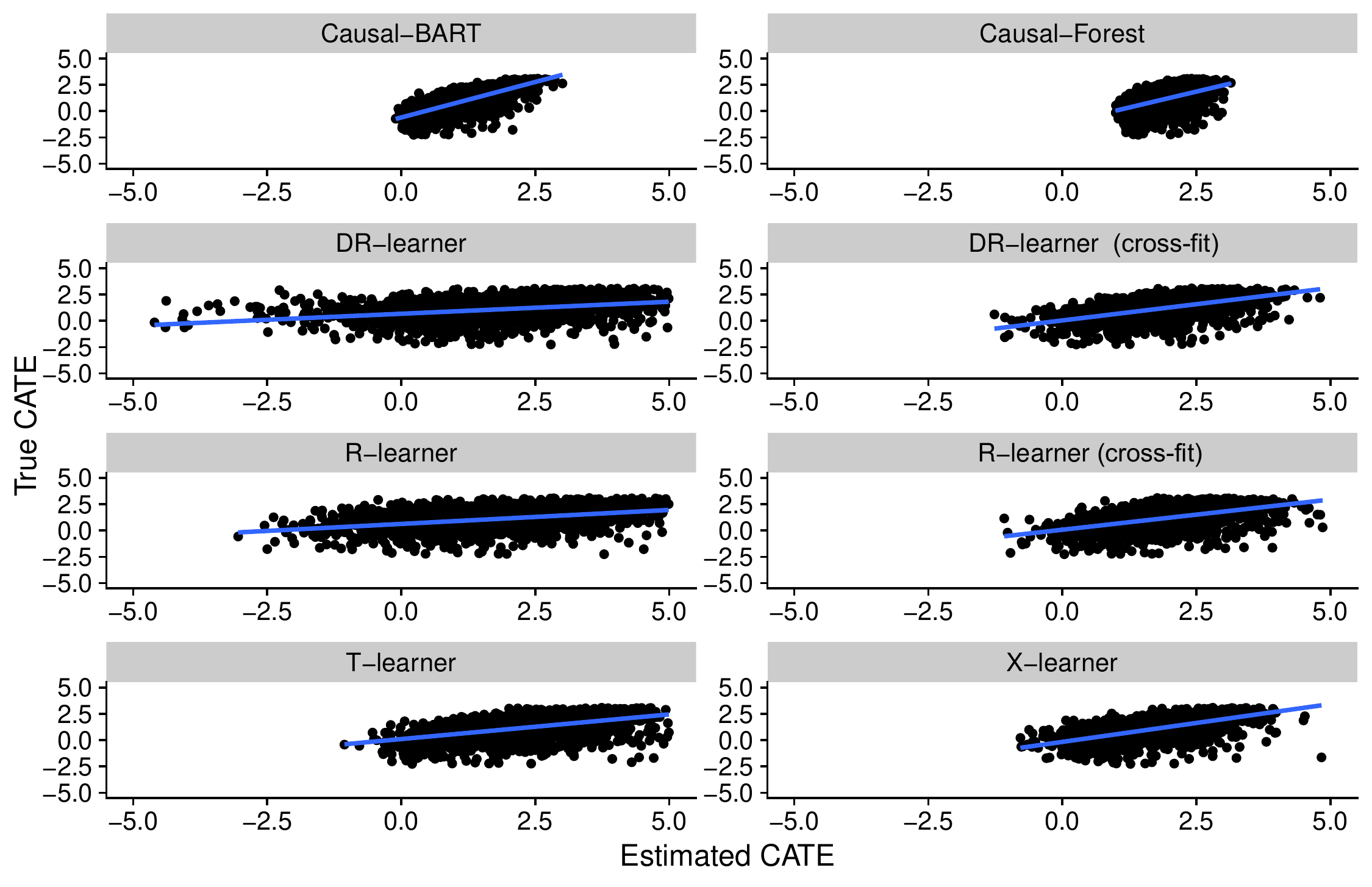}
\caption{Setting 2: Scatterplot of estimated and true CATE. \href{https://github.com/QuantLet/Meta_learner-for-Causal-ML/tree/main/Simulation-Example}{\includegraphics[width=0.02\textwidth]{qletlogo_tr.png}$_{T2}$}}
\label{fig:sim_CATE_scatter_S2}
\end{center}
\end{figure}

Figure \ref{fig:sim_CATE_sorted_S1} and \ref{fig:sim_CATE_sorted_S2} show the sorted treatment effect heterogeneity with 95\% confidence intervals for setting 1 and 2, respectively. While the causal BART method produces the lowest MSE, it has higher credible intervals than the causal forest.  Figure \ref{fig:sim_CATE_boxplot_S1} in the Appendix shows boxplots of all methods and their variation. The blue line indicates the true ATE, hence we can see how accurate all methods are to predict the ATE. We find that all methods are unbiased if the DGP is linear. The bias increases if the functions are more complex as shown in Figure \ref{fig:sim_CATE_boxplot_S2}. Figure \ref{fig:sim_CATE_boxplot_S1} and \ref{fig:sim_CATE_boxplot_S2} also shows the decrease in outliers for the DR- and R-learner when we apply additional sample splitting and cross-fitting. We do not observe these outliers for the X-learner. In Figure \ref{fig:sim_CATE_scatter_S1} and \ref{fig:sim_CATE_scatter_S2}, we plot scatterplots for the estimated vs. the true CATE. The blue line indicates a linear regression estimate with pointwise confidence intervals (around the mean) for each method. As we have seen from the MSE, the causal BART method performs best over the whole interval and in both settings. One observation is that the meta-learners estimate the CATE with higher variance (and potentially producing more outliers that need to be controlled for) than the two modified ML methods. The T-learner has the highest variance in both settings while the DR- and R-learner show the second-highest variance in setting 1 and 2. Looking at the correlation of the methods, we find that the highest correlation is between the causal BART and causal forest ($\rho = 0.85,0.81$ for the two settings). In general, the correlation is quite high in both settings (ranging from $\rho=0.64$ to $\rho=0.85$). However, we do not see any improvement in the correlation in functions that are easier to estimate.

\clearpage
\section{Conclusion}

In this tutorial, we present novel methods to estimate the conditional average treatment effect using machine learning methods. We categorize the methods into two branches. First, so-called meta-learners, that make use of off-the-shelf machine learning methods by creating a transformed outcome to estimate the CATE. They are flexible in the choice of the machine learning method as long as they converge with a specific rate. For example, we can use classification and regression trees, random forest, boosting methods, and even neural networks. The second branch contains machine learning methods that are specific designed to estimate the CATE. These methods rely on trees or an ensemble of trees like the generalized random forest, causal boosting, and a Bayesian approach using additive regression trees. The use of meta-learners needs special care because they are quite flexible in the choice of the ML method and also concerning sample splitting. We, therefore, provide pseudo-code along with R-code for many of such meta-learners and show how they can be used to estimate the CATE on the whole dataset. We also demonstrate how to use the second branch of methods by integrating the packages in R-code that uses the same data structure as the meta-learners. When possible we apply cross-fitting as an averaging procedure of a subset of the data conditional on different training folds. 

To demonstrate the strength and differences of all the methods that we consider, we present four examples. Two empirical examples, the first from a randomized control trial and the second from an observational study. The third and fourth examples contain simulated data where the true treatment effect can be observed and hence compared with the estimates from all the methods. In the empirical examples, we find strong evidence of positive treatment effects for each observation while significant heterogeneity in the effects is not that clear. This is mainly if we base the conclusion on calculated confidence intervals via the bootstrap or credible intervals. We do, however, find differences in the width of the confidence intervals and also in the CATE prediction among the methods. These differences also occur in the simulated data. We, therefore, recommend that practitioners not rely on only one method but rather use multiple methods and compare the results. One should also carefully think about the different tuning parameters that can be set when using machine learning methods. Depending on the method there can be a variety of options to consider. We try to avoid the problem of manually selecting such parameters through cross-validation and the selection of different ML methods for each nuisance function. Sample splitting and cross-fitting is a further necessary step to get robust and accurate estimates among the methods. One observation from this simulation is clearly that the meta-learners can improve in terms of MSE with simpler functions. We note that the results heavily depend on the chosen ML methods. Through applying different ML methods in the Super Learner we find that the selection of the best ML method depends on the data generating process and varies across the functions. For example, if the data structure is quite complicated and non-linear, a model based on the lasso might not be the best choice. Including more ML methods could improve the prediction accuracy depending on the data generating process. Using two-step sample splitting with cross-fitting further improves the prediction.


\clearpage
\FloatBarrier 
\newpage

\phantomsection \addcontentsline{toc}{section}{References}
\bibliographystyle{plainnat}
\bibliography{references}


\clearpage

\appendix
\section{Additional Proofs}

Proof of the doubly-robustness property for the DR-learner.  If either the propensity score or the conditional mean is correctly specified, the doubly-robust estimator is an unbiased estimator. Let us look at $\hat{mu}_1(x)$, the same procedure holds analog or $\hat{\mu}_0$. That is, $\hat{\mu}_1(x)$ estimates the following: 
\begin{align}
=&\textsf{E}\left[{\mu}_1(x)+\dfrac{D\left\{Y-{\mu}_{1}\left(x \right)\right\}}{{e}\left(x\right)} \right] \nonumber \\
= &\textsf{E}\left[\frac{e(x)}{e(x)}{\mu}_1(x) +\dfrac{D\left\{Y-{\mu}_{1}\left(x \right)\right\}}{{e}\left(x\right)} \right] \nonumber\\
= &\textsf{E}\left[\frac{DY}{e(x)} - \frac{(D-e(x))\mu_1(x)}{e(x)} \right], \text{add } \frac{e(x)}{e(x)}Y, -\frac{e(x)}{e(x)}Y \nonumber\\
 =&\textsf{E}\left[\frac{(D-e(x))Y}{e(x)} - \frac{(D-e(x))\mu_1(x)}{e(x)} +\frac{e(x)}{e(x)}Y  \right] \nonumber \\
 \hat{\mu}_1(x) = &\textsf{E}\left[\frac{(D-e(x))(Y^1-\mu_1(x))}{e(x)}  +Y^1   \right] \nonumber \\
\hat{\mu}_1(x) = &\textsf{E}[Y^1] + \textsf{E}\left[\frac{(D-e(x))(Y^1-\mu_1(x))}{e(x)}\right]	\label{equ:dr-proof}
\end{align}

\section{Additional Plots}

\begin{figure}[ht]
\begin{center}
\includegraphics[width=0.8\textwidth]{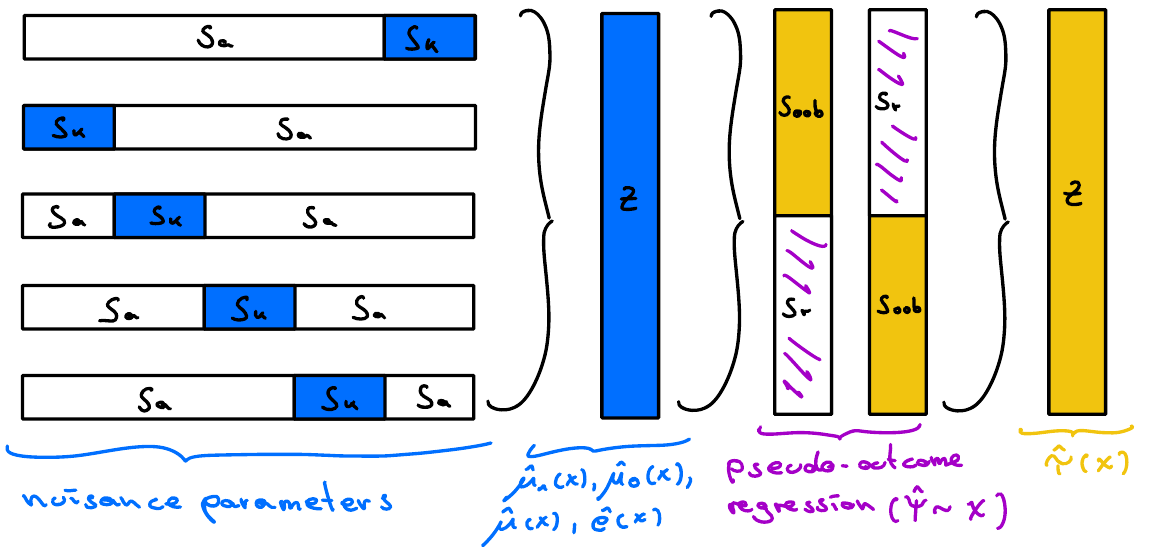}
\caption{Two-step sample splitting: 50:50, no cross-fitting.}
\label{fig:two-step_50_50}
\end{center}
\end{figure}

\begin{figure}[ht]
\begin{center}
\includegraphics[width=0.8\textwidth]{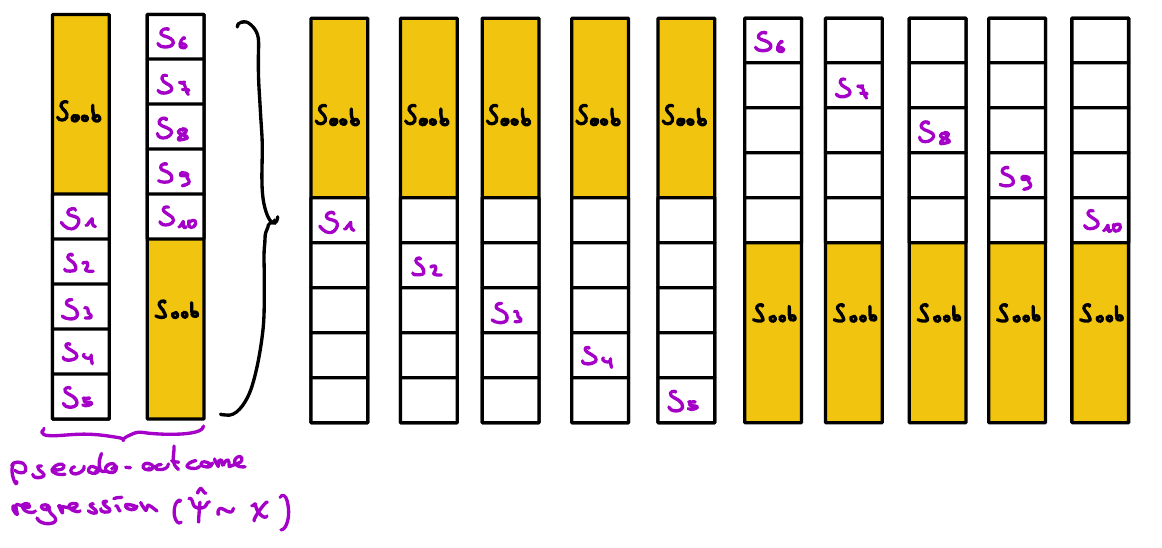}
\caption{Detailed cross-fitting procedure for 5 folds.}
\label{fig:cross-fit_detailed}
\end{center}
\end{figure}

Figure \ref{fig:two-step_50_50} describes the two-step sample splitting. The first splitting is necessary for nuisance parameter estimation. The second split is done to get out-of-bag estimates of the CATE. In this approach no cross-fitting is applied. Figure \ref{fig:cross-fit_detailed} now uses the samples that include the pseudo-outcomes and split it into different folds. Each fold is used to train a regression model. Prediction is done in the out-of-bag fold. Through the different folds we get as many predictions as folds used for training. These estimates are then averaged which applies the cross-fitting procedure. 

\vskip 0.5cm

\vskip 0.5cm

\begin{figure}[ht]
\begin{center}
\includegraphics[width=0.7\textwidth]{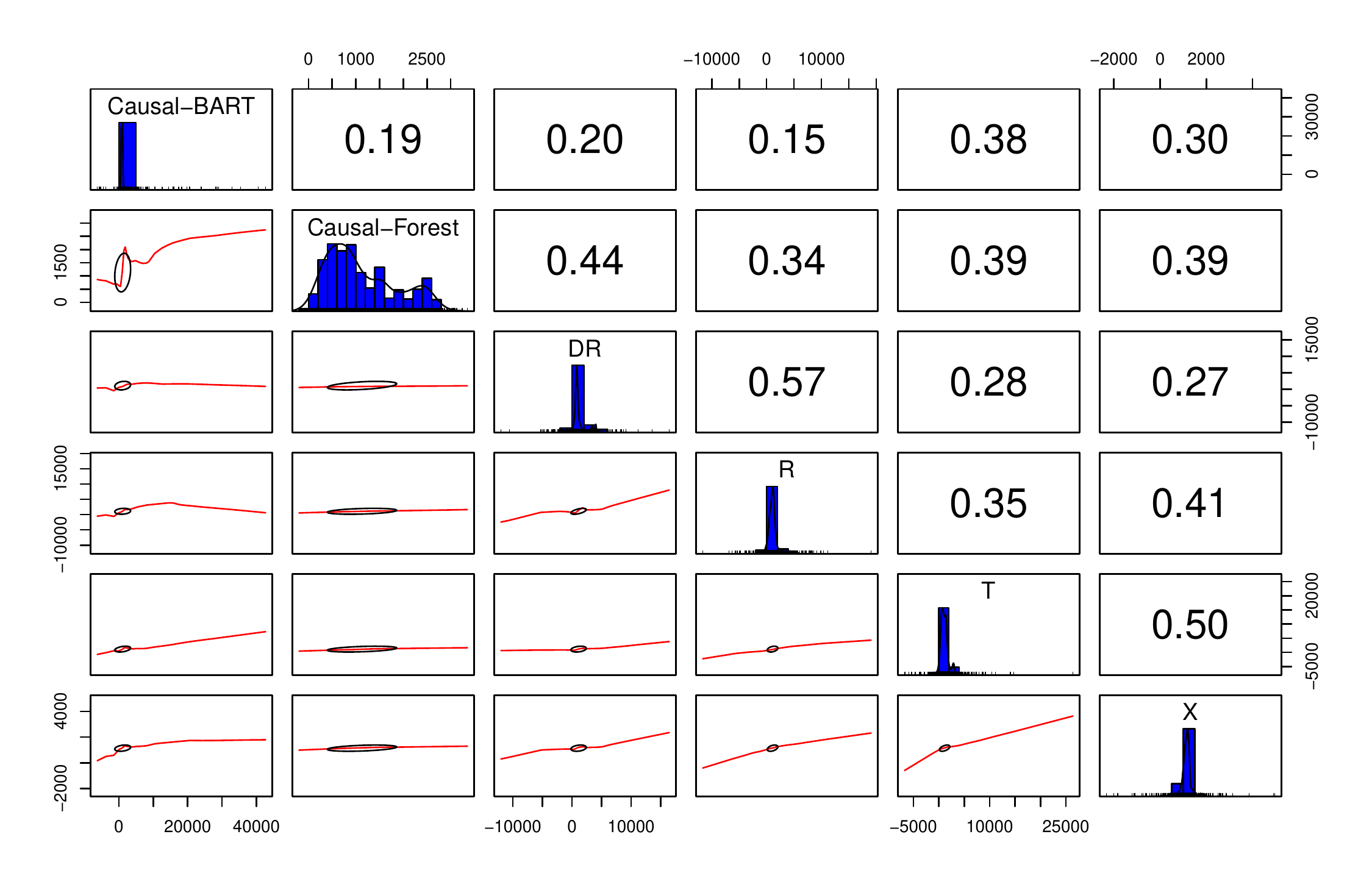}
\caption{Correlation of CATE between different methods from the microcredit data. \href{https://github.com/QuantLet/Meta_learner-for-Causal-ML/tree/main/Microcredit-Example}{\includegraphics[width=0.02\textwidth]{qletlogo_tr.png}$_{C1}$}}
\label{fig:Microcredit_corrplot}
\end{center}
\end{figure}

\vskip -1.0cm
\begin{figure}[ht]
\begin{center}
\includegraphics[width=0.7\textwidth]{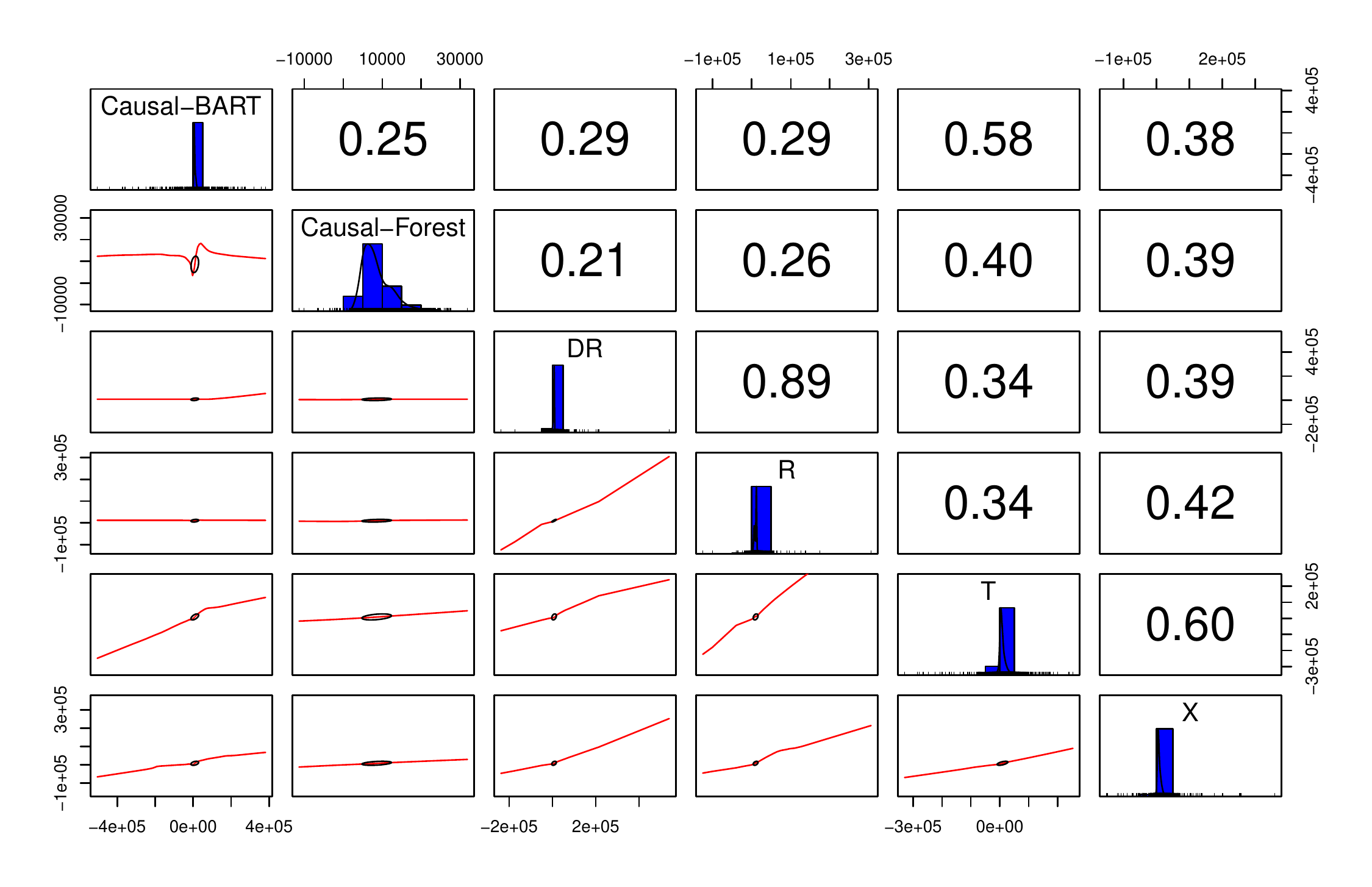}
\caption{Correlation of CATE between different methods from the 401k data. \href{https://github.com/QuantLet/Meta_learner-for-Causal-ML/tree/main/401k-Example}{\includegraphics[width=0.02\textwidth]{qletlogo_tr.png}$_{C2}$}}
\label{fig:401k_corrplot}
\end{center}
\end{figure}

\begin{figure}[ht]
\begin{center}
\includegraphics[width=0.7\textwidth]{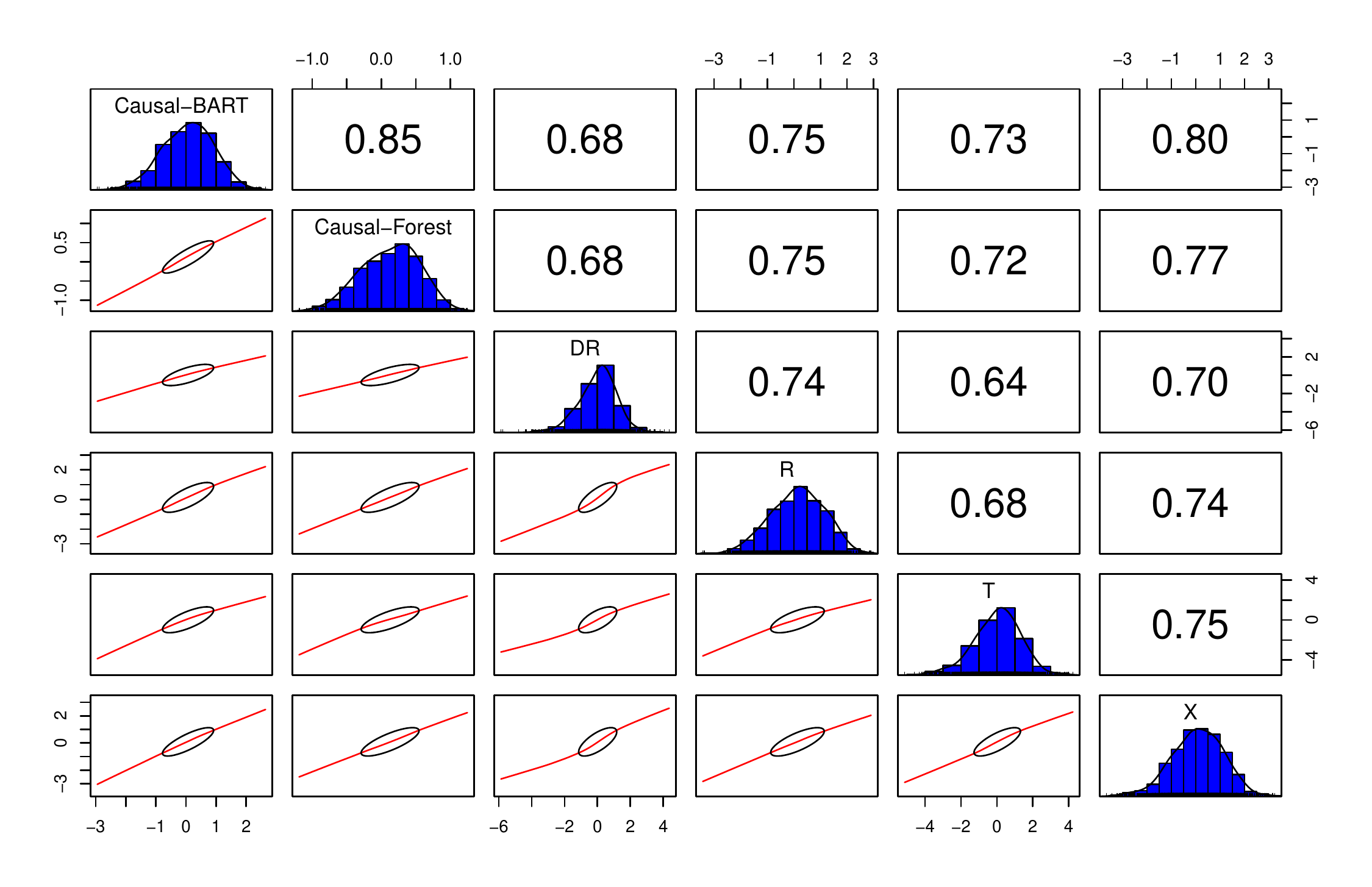}
\caption{Correlation of CATE between different methods for simulation setting 1. \href{https://github.com/QuantLet/Meta_learner-for-Causal-ML/tree/main/Simulation-Example}{\includegraphics[width=0.02\textwidth]{qletlogo_tr.png}$_{C3}$}}
\label{fig:S1_corrplot}
\end{center}
\end{figure}

\begin{figure}[ht]
\begin{center}
\includegraphics[width=0.7\textwidth]{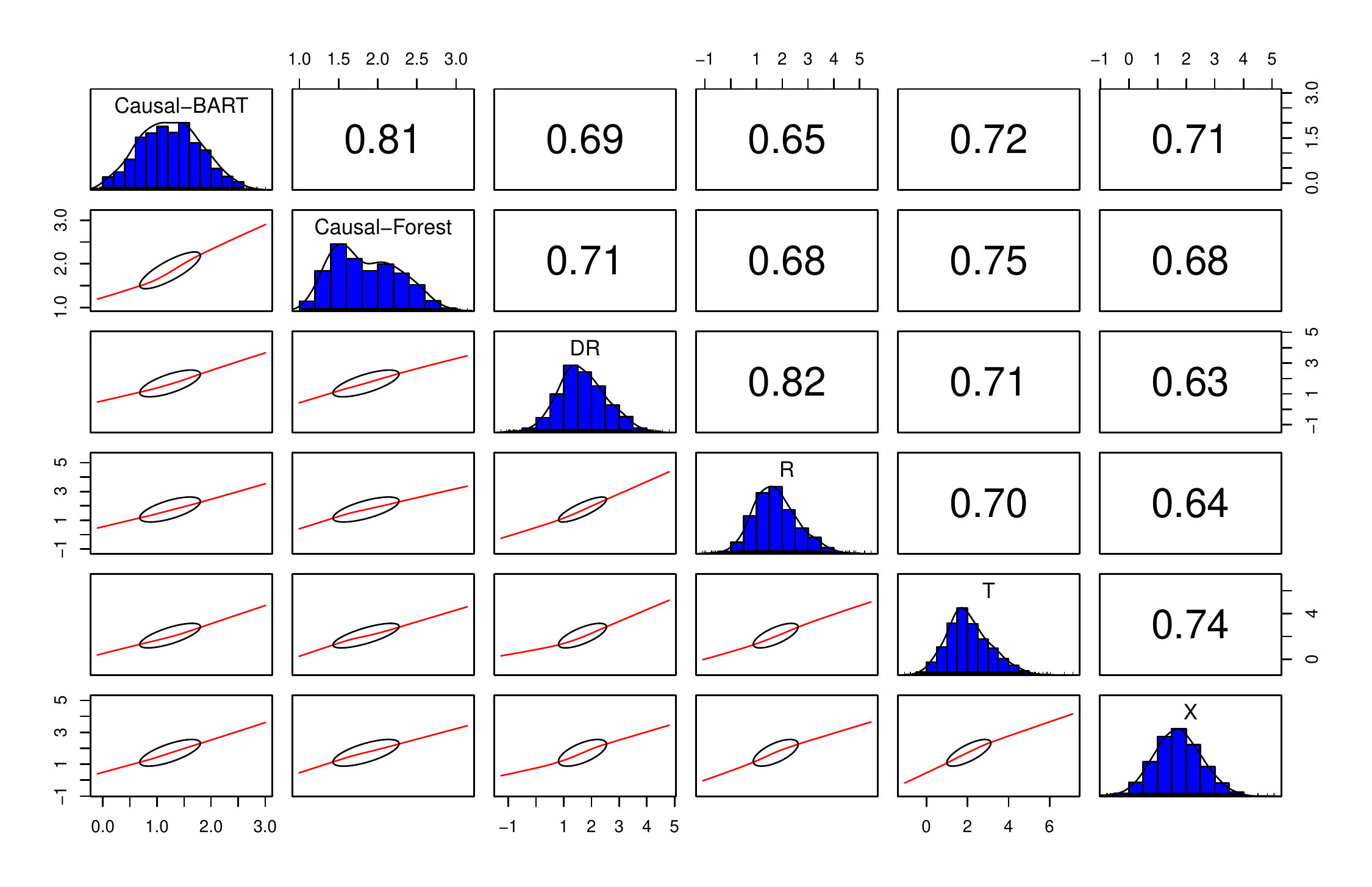}
\caption{Correlation of CATE between different methods for simulation setting 2. \href{https://github.com/QuantLet/Meta_learner-for-Causal-ML/tree/main/Simulation-Example}{\includegraphics[width=0.02\textwidth]{qletlogo_tr.png}$_{C4}$}}
\label{fig:S2_corrplot}
\end{center}
\end{figure}

\begin{figure}[ht]
\begin{center}
\includegraphics[width=0.8\textwidth]{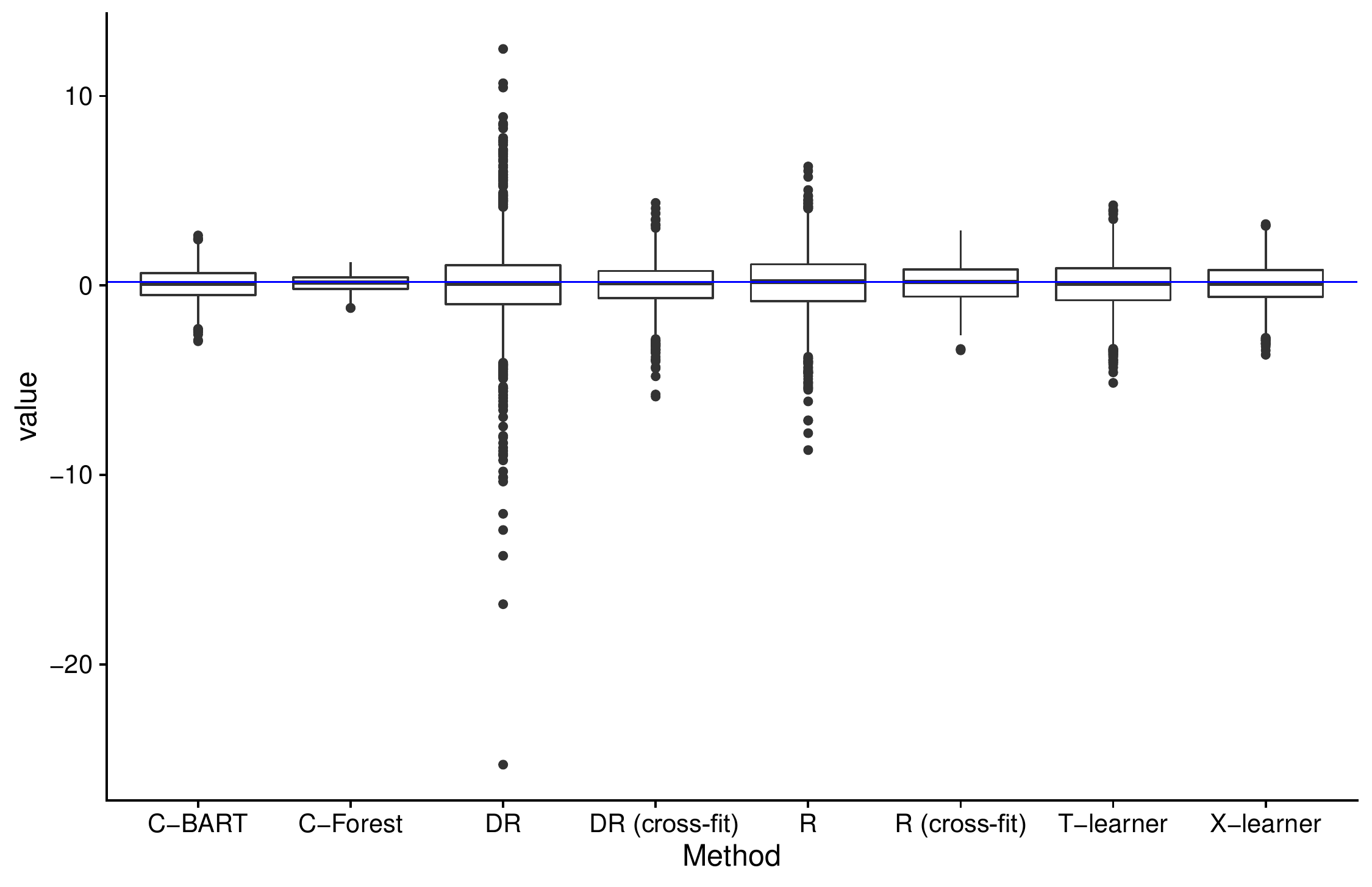}
\caption{Setting 1: Boxplots of different methods. Blue line shows true ATE. \href{https://github.com/QuantLet/Meta_learner-for-Causal-ML/tree/main/Simulation-Example}{\includegraphics[width=0.02\textwidth]{qletlogo_tr.png}$_{B3}$}}
\label{fig:sim_CATE_boxplot_S1}
\end{center}
\end{figure}

\begin{figure}[ht]
\begin{center}
\includegraphics[width=0.8\textwidth]{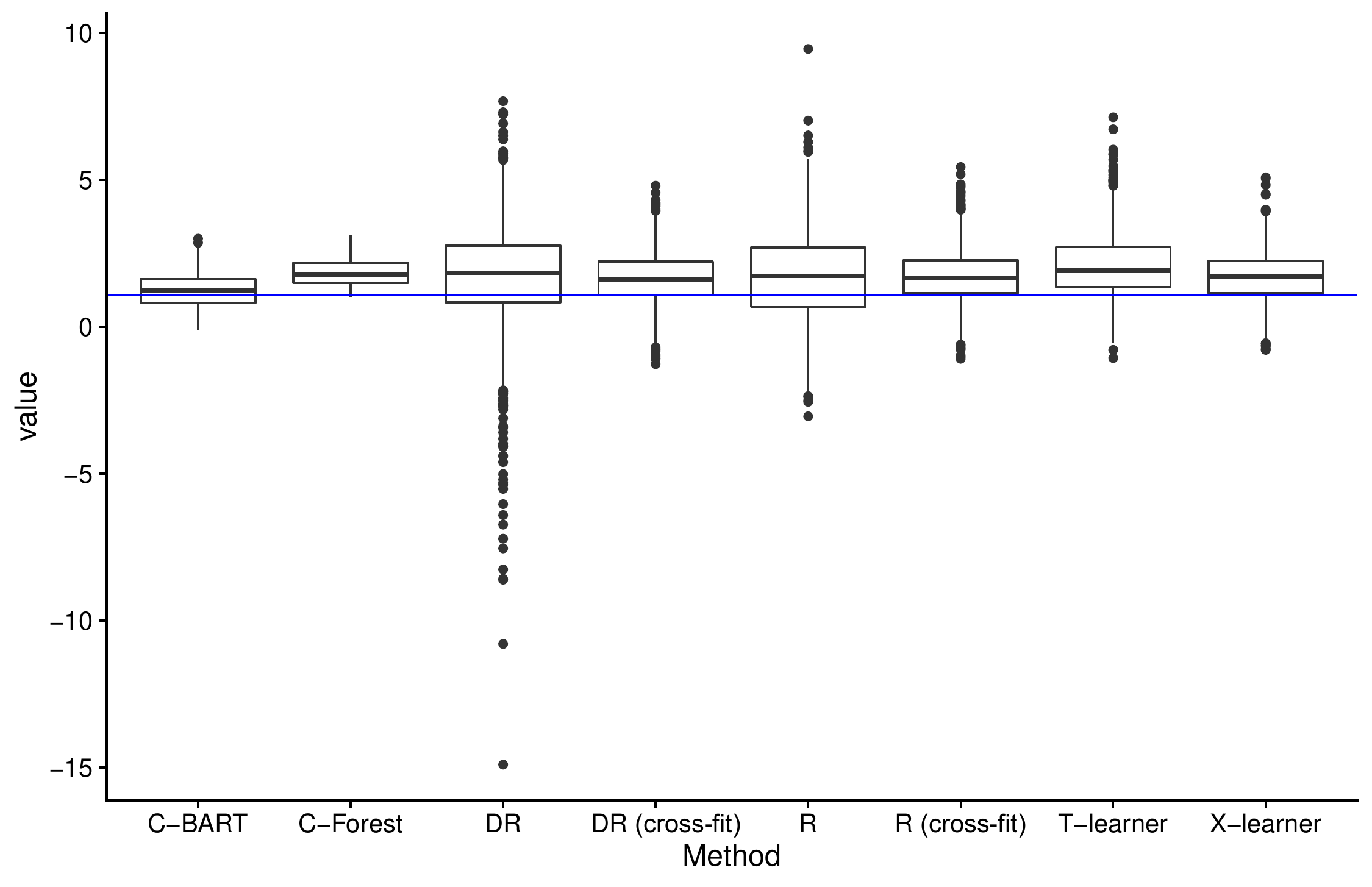}
\caption{Setting 2: Boxplots of different methods. Blue line shows true ATE. \href{https://github.com/QuantLet/Meta_learner-for-Causal-ML/tree/main/Simulation-Example}{\includegraphics[width=0.02\textwidth]{qletlogo_tr.png}$_{B4}$}}
\label{fig:sim_CATE_boxplot_S2}
\end{center}
\end{figure}

\clearpage

\clearpage

\section{Tables}

\subsection{Classification results for the microcredit example}

\begin{table}[ht]
\centering
{\renewcommand{\arraystretch}{1.5}%
\resizebox{0.9\textwidth}{!}{%
    \begin{threeparttable}
    \caption{CLAN results for different methods.}
\label{tab:micro_CLAN_full}
\begin{tabular}{llllllllll}
\hline \hline

                                                                         & \multicolumn{3}{c}{DR-learner}                   & \multicolumn{3}{c}{R-learner}                    & \multicolumn{3}{c}{T-learner}                     \\ \hline
                                                                         & Most Affected & Least Affected & Difference      & Most Affected & Least Affected & Difference      & Most Affected  & Least Affected & Difference      \\
\multirow[t]{3}{*}{Head age}                                                                   & 19.19 & 46.92 & -27.73 & 29.99 & 42.52 & -12.53 & 23.43 & 38.17 & -14.74           \\
                                                                         & (17.81,20.58) & (45.53,48.30) & (-29.68,-25.77) & (28.50,31.49) & (40.98,44.06) & (-14.67,-10.38) & (21.93,24.94) & (36.67,39.68) & (-16.87,-12.61) \\  \\
                                                                         & -             & -              & {[}0.000{]}     & -             & -              & {[}0.000{]}     & -              & -              & {[}0.000{]}     \\
\multirow[t]{3}{*}{\begin{tabular}[t]{@{}l@{}}Non-agricultural\\ self-employed\end{tabular}}  & 0.118 & 0.186 & -0.068 & 0.178 & 0.097 &  0.081 & 0.149 & 0.126 &  0.023 \\ 
   & (0.097,0.139) & (0.165,0.207) & (-0.098,-0.038) & (0.158,0.198) & (0.076,0.118) & (0.052,0.110) & (0.128,0.169) & (0.106,0.146) & (-0.006,0.051) \\ 
                                                                         & -             & -              & {[}0.000{]}     & -             & -              & {[}0.000{]}     & -              & -              & {[}0.244{]}     \\
\multirow[t]{3}{*}{\begin{tabular}[t]{@{}l@{}}Borrowed from \\ any source\end{tabular}}      & 0.138 & 0.338 & -0.201 & 0.181 & 0.320 & -0.139 & 0.116 & 0.273 & -0.157 \\ 
   & (0.113,0.162) & (0.314,0.363) & (-0.235,-0.166) & (0.155,0.206) & (0.294,0.346) & (-0.175,-0.103) & (0.093,0.139) & (0.250,0.296) & (-0.189,-0.125) \\ 
                                                                         & -             & -              & {[}0.000{]}     & -             & -              & {[}0.350{]}     & -              & -              & {[}0.000{]}     \\ \hline
                                                                         & \multicolumn{3}{c}{X-learner}                    & \multicolumn{3}{c}{Causal BART}                  & \multicolumn{3}{c}{Causal Forest}                 \\ \hline
                                                                         & Most Affected & Least Affected & Difference      & Most Affected & Least Affected & Difference      & Most Affected  & Least Affected & Difference      \\
\multirow[t]{3}{*}{Head age}                                                                 & 19.66 & 38.94 & -19.28 & 14.85 & 50.05 & -35.21 & 10.25 & 46.80 & -36.55 \\ 
   & (18.19,21.14) & (37.47,40.42) & (-21.37,-17.19) & (13.68,16.02) & (48.88,51.22) & (-36.86,-33.55) & (9.271,11.22) & (45.82,47.77) & (-37.93,-35.17) \\
                                                                         & -             & -              & {[}0.000{]}     & -             & -              & {[}0.000{]}     & -              & -              & {[}0.000{]}     \\
\multirow[t]{3}{*}{\begin{tabular}[t]{@{}l@{}}Non-agricultural\\ self-employed\end{tabular}}  & 0.138 & 0.181 & -0.043 & 0.150 & 0.099 & 0.051 & 0.073 & 0.136 & -0.064 \\ 
   & (0.116,0.160) & (0.159,0.202) & (-0.073,-0.012)  & (-0.073,-0.012) & (0.080,0.119) & (0.130,0.169) & (0.118,0.154) & (0.055,0.091) & (0.038,0.089) \\
                                                                         & -             & -              & {[}0.013{]}     & -             & -              & {[}0.000{]}     & -              & -              & {[}0.000{]}     \\
\multirow[t]{3}{*}{\begin{tabular}[t]{@{}l@{}}Borrowed from \\ any source\end{tabular}}       & 0.091 & 0.417 & -0.327 & 0.091 & 0.300 & -0.210 & 0.050 & 0.388 & -0.338 \\ 
   & (0.067,0.115) & (0.394,0.441) & (-0.360,-0.293) & (0.068,0.113) & (0.278,0.323) & (-0.242,-0.178) & (0.028,0.072) & (0.366,0.411) & (-0.370,-0.307) \\
                                                                         & -             & -              & {[}0.000{]}     & -             & -              & {[}0.000{]}     & -              & -              & {[}0.000{]}     \\ \hline \hline
\end{tabular}
\begin{tablenotes}
      \small
      \item \textit{Notes: Averages are taken from the mean of the CATE over 500 bootstrap iterations.} 
    \end{tablenotes}
  \end{threeparttable}
}
}
\end{table}

\clearpage
\section*{Additional Pseudocode}

\begin{algorithm}[ht!]
\SetKwInOut{Input}{Input}
\Input{$Z_i = \{Y_i,D_i,X_i\}_{i \in N}$} 
Split sample $Z$ into $K$ random subsets \\
\For{k in \{1, \ldots,K\}}{
\textbf{assign} Sample $S_a = Z \cupdot S_k$ and $S_k$ \\
 \textbf{regress} $Y_{i}=\hat{\mu}\left(X_{i},D_i\right)+\hat{U}_{i},$ with $i \in S_a $ \\
 \hskip 1.0cm 	\textbf{estimate} $\hat{Y}^0_{i}=\hat{\mu}\left(X_{i},D=0\right)$, with $i \in S_k$ \\
 \hskip 1.0cm 	\textbf{estimate} $\hat{Y}^1_{i}=\hat{\mu}\left(X_{i},D=1\right)$, with $i \in S_k$ \\
 \textbf{create} $\hat{\tau}_k(X_i) = \hat{\mu}(X_i,D=1) - \hat{\mu}(X_i,D=0)$ \\
  }
  \textbf{combine} $\hat{\tau}(X_i) = \{\hat{\tau}_1,\hat{\tau}_k, \ldots,\hat{\tau}_K$\} \\
  \caption{S-learner}
   \label{pseudo:s}
\end{algorithm}
\noindent

\begin{algorithm}[ht!]

\SetKwInOut{Input}{Input}
\Input{$Z_i = \{Y_i,D_i,X_i\}_{i \in N}$} 
$p:$ evaluation point (out-of-sample)\\

$S_{0}=\left\{i: D_{i}=0\right\} $\\
$S_{1}=\left\{i: D_{i}=1\right\}$ \\
$n_{0}=\# S_{0} $\\
$n_{1}=\# S_{1}$ \\

\For {$b$ in $\{1, \ldots, B\}$} { 
$S_{b}^{*}=c\left(\operatorname{sample}(S_{0}, S_1)\right)$\\
$y_{b}^{*}=y\left[s_{b}^{*}\right]$ \\
$d_{b}^{*}=d\left[s_{b}^{*}\right]$ \\
$x_{b}^{*}=x\left[s_{b}^{*}\right]$ \\

$\hat{\tau}_{b}^{*}(p)=\operatorname{learner}\left(y_{b}^{*}, d_{b}^{*}, x_{b}^{*}\right)(p)$\\
}
$\hat{\tau}(p)=\operatorname{learner}(y,d,x)(p)$ \\
$\hat{\sigma}=\operatorname{sd}\left(\left\{\hat{\tau}_{b}^{*}(p)\right\}_{b=1}^{B}\right)$ \\
$\operatorname{return}\left(\hat{\tau}(p)-q_{\alpha / 2} \hat{\sigma}, \hat{\tau}(p)+q_{1-\alpha / 2} \hat{\sigma}\right)$
\caption{Bootstrap Confidence Interval}
\end{algorithm}
\vskip 0.5cm

Algorithm \ref{pseudo:IPW} refers to the inverse probability weighting (IPW) estimator based on \cite{horvitz1952generalization}. While \cite{kuenzel2019meta} refer to this estimator as the F-learner, it is also known as the (simplest) transformed outcome estimator. This is because $\hat{\psi}_{IPW}$ is an unbiased estimate of the ATE. The only nuisance function that is needed to create this outcome is the propensity score. We then map the covariates onto the transformed outcome (or pseudo-outcome). The reason for this mapping is again to smooth the function since the IPW estimator can suffer from high variance if the propensity score estimates are near zero or one. Below we show that the IPW estimator is an unbiased estimator for the ATE. 

\begin{align*}
\mathsf{E}[\psi_{IPW}|X_i=x] &= \mathsf{E}[Y\{\frac{D}{{e}(x)} - \frac{1-D}{1-{e}\left(X_{i}\right)}\}|X_i=x] \\
&= \mathsf{E}[\{D\frac{Y}{{e}(x)} - (1-D)\frac{Y}{1-{e}\left(x\right)}\}|X_i=x] \\
&= \mathsf{P}(D=1|X_i=x)\frac{1}{e(x)} \mathsf{E}[Y|D=1,X_i=x]  \\  &\quad \quad -\mathsf{P}(D=0|X_i=x)\frac{1}{1-e(x)} \mathsf{E}[Y|D=0,X_i=x] \\
&=  \mathsf{E}[Y|D=1,X_i=x] -  \mathsf{E}[Y|D=0,X_i=x] \\
&= \mu_1(x) - \mu_0(x) = \tau(x)
\end{align*}

\begin{algorithm}[ht!] 
\SetKwInOut{Input}{Input}
\Input{$Z_i = \{Y_i,D_i,X_i\}_{i \in N}$} 
Split sample $Z$ into $K$ random subsets \\
\For{k in \{1, \ldots,K\}}{
\textbf{assign} Sample $S_a = Z \cupdot S_k$ and $S_k$ \\
\textbf{regress} $D_{i}=\hat{e}\left(X_{i}\right)+\hat{V}_{i},$ with $i \in S_a $ \\
 \hskip 1.0cm 	\textbf{estimate} $\hat{D}_{i}=\hat{e}\left(X_{i}\right)$, with $i \in S_k$ \\
\textbf{create} $\hat{\psi}_{IPW} = Y \{\frac{D}{\hat{e}(x)} - \frac{1-D}{1-\hat{e}\left(X_{i}\right)}\}$ \label{pseudo:IPW:TO}\\
 \textbf{store} $\hat{\psi}_{IPW,k}$ for $i \in S_k$ \\
}
 \textit{Cross-fitting:}\\
 \For{oob in (1:2)}{
 \textbf{if} oob = 1: $S_{oob} = Z_i$ with $i \in \{1,...,N/2\}$ and $S_{train} = Z_i \cupdot S_{oob}$ \\
 \textbf{if} oob = 2: $S_{train} = Z_i$ with $i \in \{1,...,N/2\}$ and $S_{oob} = Z_i\cupdot S_{in}$ \\
 \For{l in 1:5}{
 \textbf{split} $S_{train}$ in $\{S_1,S_2,\ldots,S_5\}$ \\
 \textbf{regress} $\hat{\psi}_{i} = \hat{t}_{IPW}(X_i) + W_i$, for $i \in S_{l}$ \\
  \hskip 1.0cm\textbf{estimate} $\tilde{\tau}_{l}(X_i) = \hat{t}_{IPW}(X_i)$, with $i \in S_{oob}$ \\
  }
  \textbf{average} $\hat{\tau}_{oob}(X_i) = \textsf{E}[\tilde{\tau}(X_i)]$\\ 
  }
 \textbf{row bind} $\hat{\tau}(X_i) = \{\hat{\tau}_1,\hat{\tau}_2\}$ \\
  \caption{IPW-learner}
  \label{pseudo:IPW}
\end{algorithm}
\noindent
\vskip 0.5cm

\end{document}